%% file: main.tex
\documentclass[acmlarge]{acmart}

\AtBeginDocument{%
  \providecommand\BibTeX{{%
    \normalfont B\kern-0.5em{\scshape i\kern-0.25em b}\kern-0.8em\TeX}}}

\acmBooktitle{Survey of Disaggregated Memory}

\renewcommand\footnotetextcopyrightpermission[1]{} % removes footnote with conference information in first column
\usepackage{hyperref}
\usepackage{tabularx}
\usepackage{subfig,graphicx}
\usepackage {color} 
\usepackage{multirow}
\usepackage{makecell}

\begin{document}

%\title{Survey of Disaggregated Memory: Techniques at Architecture, System, and Application Level }
%\title{Survey of Disaggregated Memory: System Design, Optimization Technique, and Future Opportunity}
% \title{Survey of Disaggregated Memory: Key Technical Foundation of Elastic and Efficient Next-Generation Datacenter}
% \title{Survey of Disaggregated Memory: Technique  across Architecture, System, and Application Layers for Next-Generation Elastic Datacenters}
% \title{Survey of Disaggregated Memory: Technique Insights across Architecture, System, and Application for Next-generation Datacenters}
\title{Survey of Disaggregated Memory: Cross-layer Technique Insights for Next-Generation Datacenters}
% \subtitle{What we talk about when we talk about disaggregated memory, i.e. far memory}
\renewcommand\shorttitle{Survey of Disaggregated Memory}
% \author{xxx}
% % \authornote{Both authors contributed equally to this research.}
% \affiliation{%
%   \institution{Shanghai Jiao Tong University, China}
% }

\author{Jing Wang, Chao Li*, Taolei Wang, Jinyang Guo, Hanzhang Yang, Yiming Zhuansun, Minyi Guo}
\email{{jing618,sjtuwtl,lazarus,linqinluli,zsym2019}@sjtu.edu.cn, {lichao, guo-my}@cs.sjtu.edu.cn}
\affiliation{%
  \institution{Shanghai Jiao Tong University}
  \city{Shanghai}
  \country{China}
}

\renewcommand{\shortauthors}{Jing Wang, et al.}
\begin{abstract}
The growing scale of data requires efficient memory subsystems with large memory capacity and high memory performance. Disaggregated architecture has become a promising solution for today's cloud and edge computing for its scalability and elasticity. As a critical part of disaggregation, disaggregated memory faces many design challenges in many dimensions, including hardware scalability, architecture structure, software system design, application programmability, resource allocation, power management, etc. These challenges inspire a number of novel solutions at different system levels to improve system efficiency. In this paper, we provide a comprehensive review of disaggregated memory, including the methodology and technologies of disaggregated memory system foundation, optimization, and management. We study the technical essentials of disaggregated memory systems and analyze them from the hardware, architecture, system, and application levels. Then, we compare the design details of typical cross-layer designs on disaggregated memory. Finally, we discuss the challenges and opportunities of future disaggregated memory works that serve better for next-generation elastic and efficient datacenters.
\end{abstract}





\maketitle
\section{Introduction}
%应用需求驱动分离式内存架构的提出
The scale of data-driven intelligent services today has grown exponentially, including video processing, natural language processing, graph processing, etc. The large amount of data used for applications provides better quality of analysis results, which requires careful memory resource management. Nevertheless, this aggravates the problem of extremely insufficient memory resources. Handling in-memory data poses great changes to today's hardware and software system designs. Both academic research and industrial deployment require efficient subsystems with larger memory capacity, higher computation performance, and higher resource utilization.  

%these computation-oriented and storage-oriented 
%the demand for memory-intensive workloads, such as graph processing, machine learning, and data mining, has experienced rapid growth.  
%Most systems today face memory bottleneck problem that blocks performance improvement of memory and storage devices with different latency and bandwidth.
%分离式内存架构的硬件基础
% 
% Traditional data centers consisting of monolithic servers are reaching their limits in terms of scalability, performance, and efficiency. 
Most data center environments face memory bottleneck problems that block performance improvement of memory and storage devices. For example, hardware limitations necessitate the adoption of distributed computing and out-of-core processing in applications. One has to divide data into smaller parts or compress data size to handle memory shortage. In addition, today's cloud applications often rely on diverse memory-intensive services and runtimes, occupying nearly 70\% of the workload  \cite{serverless-percent}. The memory-related resource consumption/overhead of these services is significantly growing, which is known as the \textit{memory tax} \cite{tmo} phenomenon. 

% Due to hardware limitations, applications have to adopt distributed computing and out-of-core processing to divide data into several parts, or compress the data size to fit into the limited memory, which is named as \textit{memory constraint} phenomenon. 
%which are caused by \textit{memory constraints} and \textit{memory tax} status.

To address the above issues, one solution is to implement a large disaggregated memory pool and provide an unlimited memory resource, thereby alleviating the resource pressure. \textit{Disaggregated architecture} is proposed as a flexible and composable infrastructure  \cite{cdi,disaggregated-mem}, breaking the limits of scalability, performance, and efficiency on classic monolithic servers. Disaggregation allows applications to use any required ratio of resources of computing units, memory, storage, networks, accelerators, etc. One can re-aggregate disaggregated resources to compose entities and build upper-layer resource abstraction to utilize it efficiently and flexibly. Disaggregation can also solve resource fragmentation problems due to unbalanced resource scheduling on each machine. 

%In disaggregated architectures, one can place more CPU cores, faster memory, and larger storage to scale up each server or add more blade servers to scale out clusters. 

%In today's data center, taking Alibaba data center as an example \cite{alianalysis}, there have been many underutilized CPU cores and memory fragments due to unbalanced resource scheduling on each machine. Disaggregation can solve resource fragmentation problems. 

%This can better utilize these resources by virtualizing and integrating the resources into a shared resource pool. 

% To reduce the resource pressure under memory constraint and memory tax problems, one solution is to build a large \textit{disaggregated} memory pool so that an unlimited memory resource can be obtained. 

%they still need efficient hardware and software codesigned environments to improve the overall computing capacity. 

\begin{figure}[t]
    \centering
    \includegraphics[width=\linewidth]{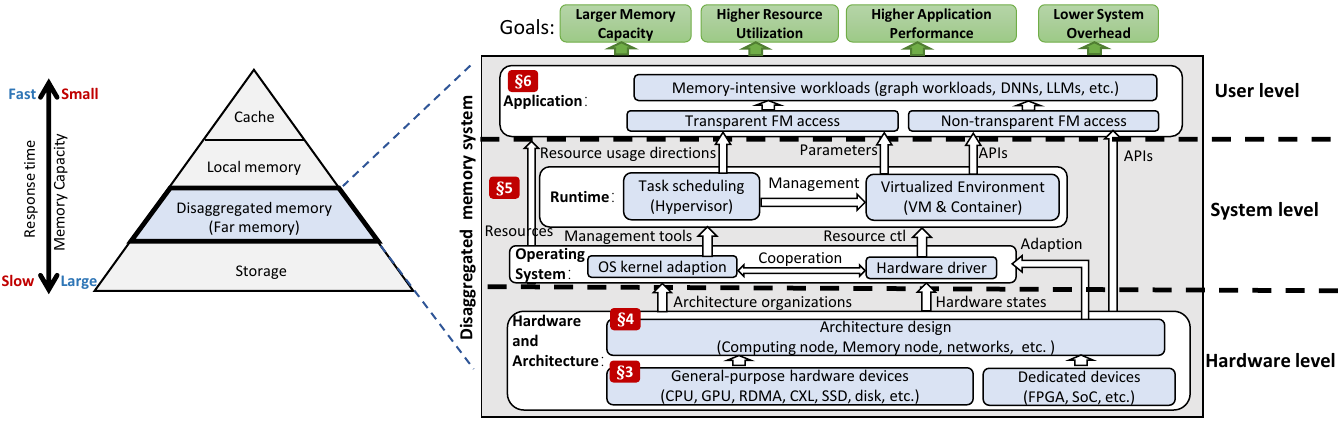}
    \caption{This survey focuses on disaggregated memory(DM), which acts as an additional layer in the existing memory hierarchy. The survey mainly studies the design goal, design level, and key design points of the DM system. }
    %需要章节对应文章内容
    \label{fig:system-levels}
    \vspace{-0.5cm}
\end{figure}

% The key design challenges is \textit{memory disaggregation}. 
%While memory disaggregation has many benefits, it also faces performance problems, especially memory performance problems, mainly from the design of \textit{disaggregated memory (DM)}. 
%compared with behaviors of I/O operations from storage devices or data communication from network devices

Specifically, \textit{disaggregated memory} (DM) aims to decouple CPU and memory resources so that one can flexibly manage these memory resources in a large and shared memory pool. Different from disaggregated storage and networks, the overhead of remote memory accesses can seriously slow down computation. To improve DM access performance, existing works try hard to adopt and design large-capacity memory devices, low-latency storage devices, and high-speed networks \cite{cxl,sumsung-ssd,nvidia-rdma}. These DM devices are seen as a larger but slower "far memory" (local memory is its counterpart) and are added into the original memory hierarchy, as Figure \ref{fig:system-levels} shows.

Advanced memory and network devices may not guarantee high memory performance if the software systems are poorly designed. Existing operating systems (OS) and runtime systems face the problem of adapting to new DM hardware. Conventional cache-based hierarchical memory management methods can not exploit the high bandwidth and large capacity of emerging devices. Thus, learning about the key technologies, optimization opportunities, and future challenges is vital when designing high-performance and high-efficiency DM systems. There is an urgent need for a comprehensive and in-depth survey to provide a full-stack description of DM systems in the hardware, system, software, and applications layers.

 This paper explores the hardware usage, architecture design, system adaption, runtime optimization, and future opportunities of DM systems. To our knowledge, it is the first survey to provide a comprehensive classification of technique optimization methods for DM systems. We discuss not only key techniques that lay the foundation for DM systems, but also the latest technologies related to memory hierarchies, heterogeneous devices, and direct-connect networks. We present the trend in heterogeneous memory design, the construction of disaggregated memory pools supporting heterogeneous environments with accelerators, as well as optimization methods for emerging AI applications. We show that, DM systems require well-designed architectural organization methods to better meet the needs of new types of applications and hardware.

 This survey is organized as follows. As shown in Figure \ref{fig:system-levels}, we first give an overview of disaggregated memory systems in Section \ref{sec-overview}. We then introduce hardware usage (Section \ref{sec-hardwareusage}), architecture design (Section \ref{sec-architecture}), OS and runtime layer development (Section \ref{sec-system}), and application-oriented optimization methods (Section \ref{sec-runtime}) of Disaggregated memory. We summarize some classic DM systems with cross-layer design and give a general analysis in Section \ref{sec-classification}. Finally, we discuss future work (Section \ref{sec-futurework}) and conclude the paper (Section \ref{sec-conclusion}).

% \begin{figure*}
%     \centering
%     \includegraphics[width=\linewidth]{figure/structure.pdf}
%     \caption{The overall structure of this survey on disaggregated memory architecture. }
%     \label{fig:structure}
% \end{figure*}

\input{Contents/2-Overview}

\input{Contents/3-Architecture}

\input{Contents/4-System-adaption}

\input{Contents/5-Application-level}

\input{Contents/7-Classification}

\input{Contents/8-Futurework}

% \begin{figure}
%     \centering
%     \includegraphics[width=0.9\linewidth]{figure/background1.pdf}
%     \caption{Different architectures for far memory platforms.}
%     \label{fig:background}
% \end{figure}

\section{Conclusion}\label{sec-conclusion}

In this paper, we summarize the concept and implementation of disaggregated memory. We have developed a detailed taxonomy of disaggregated memory systems with systematically analysis. We introduce the technique insights of disaggregated memory design and optimization at different system layers, including hardware, architecture, system, and application layers. Lastly, we delved into the potential future opportunities of disaggregated memory, aiming to offer insightful guidance for forthcoming research and developments in this field. This comprehensive review serves as a foundation for understanding on how to design disaggregated memory for the next-generation elastic and efficient datacenters.

% \bibliographystyle{plain}
\bibliographystyle{ACM-Reference-Format}

% \bibliography{thebib}
\bibliography{all-bib}
% \appendix









\end{document}

%% file: Contents/2-Overview.tex
\section{Overview} \label{sec-overview}
% Disaggregated memory in computing systems is a new paradigm where memory resources are decoupled from central processing units. Disaggregation addresses the growing gap between CPU processing capabilities and memory demands in modern data centers and large-scale computing environments.

In this section, we first introduce the related survey works on disaggregated memory. We then discuss general design considerations of DM systems, including the design goals, features, and design basis, etc.

%Finally, we give a brief overview of this paper. 
% Wang et al.  \cite{dsm-md} argue that the next wave in database system innovation should be shared-memory designs enabled by RDMA-based memory disaggregation (MD). Similar to previous design evolution in decoupling system components in favor of scalability and elasticity, we believe that MD is the key to moving database design to the next frontier. We present the challenges and opportunities in realizing distributed shared-memory databases (DSM-DB) with MD. This vision paper
% focuses on OLTP main-memory databases for high performance.

% To improve the overall computing capacity, one can place more CPU cores, faster memory, and larger storage to scale up each server or add more blade servers to scale out cluster scales. However, this is not helpful for resource usage utilization and performance, causing a large amount of cluster costs. Resource efficiency requires better design of hardware architecture with corresponding software support. Disaggregation is proposed to solve the above problem. 

% 1). memory space reclaim (garbage collection)
% 2). data separation decision
% 3). Far memory access

\subsection{Related Survey on Disaggregated Memory}
Disaggregated memory was first discussed by Lim et al.  \cite{disaggregated-mem-isca09} in 2009. They propose a general-purpose architectural building block, i.e., a memory blade, that allows memory to be "disaggregated" across a system ensemble.
 % Another widely recognized framework is the software-defined hardware infrastructure (SDHI).
%  This
% remote memory blade can be used for memory capacity expansion
% to improve performance and for sharing memory across servers to
% reduce provisioning and power costs. They use this memory blade
% building block to propose two new system architecture
% solutions—(1) page-swapped remote memory at the virtualization
% layer, and (2) block-access remote memory with support in the
% coherence hardware—that enable transparent memory expansion
% and sharing on commodity-based systems.
About a decade ago, in the 2010s, resource disaggregation was proposed. In 2016, Maciej et al. \cite{sv-bielski2016} conducted a survey on memory interconnection and sharing methods for HPC systems concerning scalability and virtualization support. Gao et al.  \cite{gao2016network} identified several research opportunities in this area, such as low-latency network, disaggregation-aware scheduler, and new programming models. Li et al.  \cite{survey-lichao} provided a survey on architecture design of edge-oriented computing paradigms, with a dedicated analysis of memory allocation.

The concept has been widely-used that disaggregation is for aggregation. Composable Disaggregated Infrastructure (CDI) \cite{cdi} is introduced and becomes well known, which maintains disaggregated memory, disaggregated storage, and disaggregated network \cite{sv-sdc} and reorganizes the necessary resources. 
 A cluster comprising multiple monolithic machines is established in this setup, underpinned by a Software-Defined Data Center (SDC).

In 2018, software-defined hardware infrastructures (SDHI) are proposed based on resource disaggregation \cite{sv-roozbeh2018}, which aims to bring greater modularity, flexibility and extensibility to cloud infrastructures. Compared with conventional server-oriented software-defined infrastructure (SDI), SDHI can decompose the network control planes \cite{sv-sdn}, storage control planes \cite{sv-sds}, software abstraction layers with computing function virtualization  \cite{sv-sdc}, etc.
Wang et al.  \cite{rdma-vldb} argued that the next wave in database system innovation should be shared-memory designs enabled by RDMA-based memory disaggregation.

% There are several survey works on the methods of optimizing memory allocation and far memory access techniques.  

The most recent surveys discuss the opportunities and challenges of memory disaggregation on advanced memory devices. Li et al. \cite{sv-clouddatabase} analyzed the taxonomy and key techniques regarding cloud-native disaggregated storage management, etc. Liu et al. \cite{icdcs19} discussed some demands and challenges of memory disaggregation in cloud data centers and examined some promising research issues to show the opportunity of memory disaggregation in virtualized environments. Ewais et al.  \cite{dmsurvey-access2023} provided a list of disaggregated memory works and discussed their basic architectures, implementations, and requirements.  Tom et al. \cite{tamingMemory} gave an introduction to CXL memory \cite{cxl}, which is a representative protocol to enable the creation of pools of memory and accelerators.

We observe that related surveys have not summarized the optimization techniques for better application performance and system efficiency on disaggregated memory. This requires a deeper analysis of the design methodology of the system.  In this survey, we take the first step to study the state-of-the-art techniques on disaggregated memory at each system level, providing a comprehensive overview of the current landscape.

\subsection{Design Considerations of Disaggregated Memory}

In this subsection, we summarize the design considerations of creating an efficient DM system for different purposes. We first introduce the design goals of DM systems and then list important features of DM devices. %We then give the arguments of definitions and terminologies in fields of disaggregated memory. 
\subsubsection{Design goals} 
A well-designed disaggregated memory system aims to combine the benefits of a large memory pool with the performance, flexibility, and reliability required to support data-hungry applications. These systems are envisioned to be particularly beneficial in environments with dynamic workloads and heterogeneous computing resources.
We summarize the design goals of the disaggregated memory system as follows. %Large Available Memory Space, High Performance of Applications, High Flexibility of Resource Usage, and High Reliability of network and memory subsystems.
\begin{itemize}
\item \textbf{Large available memory capacity: }  A primary objective is to provide a substantially larger memory pool than what is typically available in traditional monolithic servers, overcoming the physical limitations of in-server memory capacity\cite{stringfigure}. Providing ample memory space to the central processing units, including CPUs and GPUs, can bring more chances for big data computation. %This requires dynamic allocation of memory based on demand, effectively .

\item \textbf{High resource usage flexibility: }  Disaggregated memory should allow more granular and efficient allocations of memory resources, adapting to varying workload requirements. This flexibility enables better resource utilization, reduces wastage, and can save costs, especially in multi-tenant or cloud-based environments where resource demands fluctuate significantly\cite{legoos}.

\item \textbf{Reliable network and memory subsystems: } When memory resources are accessed over a network, its network architecture must be robust and fault-tolerant. Similarly, the memory subsystem must be reliable, with mechanisms to protect against data loss or corruption\cite{carbink}. This includes implementing redundancy, error detection, and correction techniques to maintain the integrity and availability of data.

\item \textbf{High application performance: }  Disaggregated memory is supposed to provide low-latency, high-throughput access to memory resources, even when these resources are not physically co-located with the CPU. The system should minimize performance overheads associated with remote memory access, ensuring that applications can leverage the large memory space without compromising speed and efficiency\cite{canvas-nsdi23}.

\end{itemize}

\subsubsection{Features of DM} 

%Such a novel system design should support virtualization techniques which are necessary to achieve together flexibility, good performance, security and fault recovery\textcolor{blue}{\{tbd?\}}.
%\subsubsection{Hardware-related design consideration}

Unlike the integrated monolithic servers in traditional architecture, research on DM architecture focuses on the novel design of \textit{computing nodes} and \textit{memory nodes}. In other words, servers will serve the applications as different roles. In the past, computing resources obtained data from memory and storage devices hierarchically, managed by the memory control unit inside each CPU. In disaggregated architecture, memory and storage resources are generally managed in pools. 
From the view of the computing units, disaggregated memory constitutes a memory entity far from local . Therefore, in most academic work, "\textit{far memory}" is often used to refer to disaggregated memory.

% \textit{Far memory: }  Compared to local memory, disaggregated memory actually constitutes a memory entity that is far than local. Therefore, in most academic work, the term "far memory" is used figuratively to refer to disaggregated memory. If an application requires a software system to access and manage the disaggregated memory device, the software system that can support far memory access is referred to as the far memory system. 
% Compare the traditional servers, the computing nodes and memory nodes have different additional functions, which is listed as follows.  

% \textbf{Additional functions of computing nodes:}
Firstly, the computing nodes should keep the following additional capabilities to support high-performance data offloading to far memory.
\begin{itemize}
\item \textbf{DM latency awareness: } Accessing a "far" memory device on the computing nodes involves longer access latency and lower bandwidth than using local DRAM\cite{disaggregated-mem-isca09}. One should leverage and amortize this overhead by carefully designing the far memory access patterns of local data processing.  

\item \textbf{Smart data offloading: } Compute nodes are responsible for the overall computational flow control of the program\cite{fastswap}. Programs often have complex iterations that control the lifecycle of copies of the required data, and these characteristics make it very difficult to offload data to far memory.

\item \textbf{Fine-grained resource orchestration: } Application utilizes multiple and heterogeneous computing, memory, network, and storage resources to complete user requests\cite{legoos}. Improving system utilization requires efficient scheduling methods to reclaim and allocate these resources well.  
\end{itemize}

Secondly, the memory nodes provide key functions to support large-capacity data storage and efficient data management strategy.
% \textbf{Additional functions of memory nodes:}
\begin{itemize}
\item \textbf{Data-related workflow control:} Memory nodes are responsible for storing the arriving data and fetching the corresponding data\cite{icde-dlsm-disaggregatedmem}. However, far memory access has higher latency compared with a storage system. Thus, users must be careful with each step of the data flow process. 

\item \textbf{High-performance far memory access:} DM serves as near-computing memory devices that show much higher performance than traditional data providers. For example, solid state disks (SSD) with much higher bandwidth and IOPS than disks can be recognized as a far memory space\cite{tmo}. 

\item \textbf{Heterogeneous memory management:} Memory nodes often consist of heterogeneous memory devices on different far-memory access paths, especially when applications need to access large amounts of memory. Heterogeneous memory management is important for overall bandwidth and efficiency\cite{wang-xdm-sc24}. 
\end{itemize}

% \subsubsection{Design basis of far memory}
% The key basis of the optimization design on disaggregated memory systems is to figure out where the design located at the system level and how to access a far memory in the disaggregated memory pool.  

% A disaggregated memory system architecture should be capable of satisfying the following characteristics \cite{Carbink, Hydra} :

%\subsubsection{Software-related design consideration}

% \subsection{Design Scope of Disaggregated Memory}

\subsubsection{DM system levels} 
% The fundamental basis of the optimization design on disaggregated memory systems is at the system level. 
Generally, existing works often design DM systems at three levels.
\begin{itemize}

    \item \textbf{Architecture-level design:} One can design the DM system in a bottom-up way. The key objective is to fully utilize hardware resources in a cost-effective manner. Computing and memory nodes are expected to be connected through high-bandwidth networks. A good organization of them can provide high task performance and high data throughput. 
    
    \item \textbf{OS-level design:} With appropriate hardware support, the OS-level DM design aims to utilize hardware efficiently. Given drivers and protocols for each new-added memory and network device, one can adjust the existing operation systems to provide application-transparent DM management. One can also design virtualized DM runtimes for resource isolation and task throughput. 
    
    \item \textbf{Application-level design:} DM systems can also be designed in a top-down way. For example, one can design application-specific optimization methods that improve the efficiency of computing workflow and far memory access. Accelerators with bare metal performance also need to consider memory extension. These designs often provide non-transparent far memory access with easy-to-use and high-performance programming interfaces. 

\end{itemize}

\subsection{Paper Overview}

% We summaries the overall terminology classification related to Disaggregated Memory that occurs in this paper in Figure \ref{fig:terminology}. 

\begin{figure}[t]
    \centering
    \includegraphics[width=\linewidth]{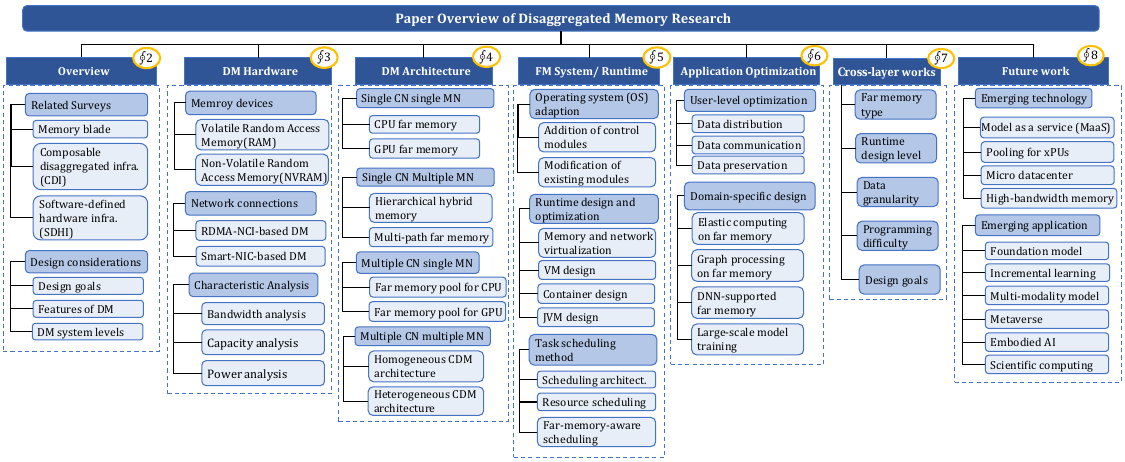}
    \caption{Overview of this survey.}
    \label{fig:terminology}
\end{figure}

We summarize the key words related to disaggregated memory in this paper, referring to Figure \ref{fig:terminology}.
% At the level of cluster infrastructure, there are some pre-defined concepts about disaggregated devices. The composable disaggregated infrastructure (CDI) \cite{cdi} maintains separated devices and reorganizes the required resources. This includes the disaggregated memory (DM), disaggregated storage (DS) and disaggregated network(DN). Another popular description is the software-defined hardware infrastructure (SDHI), which is also known as software-defined  infrastructure (SDI). The cluster  with multiple monolithic machines with  software-defined environment (SDE) building on is called Software-defined data center (SDC). 

\textbf{DM Hardware:}  Disaggregated memory requires new architecture/hardware designs for better elasticity and efficiency. Research on DM devices can be classified into three types. Memory devices usually communicate with CPUs through on-chip buses and interfaces. Generally, two types of memory devices, RAM and NVRAM, can act as disaggregated memory, with different data persistence capabilities. The network connections used for disaggregated memory (or far memory) mainly involve network interface cards (NIC) and programmable-chip-based smart NICs. In addition, hardware feature analysis is also a major part of previous work, including bandwidth, capacity and power.
%\textbf{DM Hardware:} Research on DM devices can be classified into three types. The \textit{direct-connected} DM devices can generally be plugged directly into the motherboard via appropriate slot and bus, including storage-based DMs, PCIe-based DMs, and bus-based DMs. \textit{Fabric-attached} DM can be accessed through a fabric network. It requires RDMA, smart NIC, and fabric-attached memory space with a specific memory control unit. \textit{Rack-level memory switches} for DM mainly refer to PCIe switches and user-defined programmable switches, which provide considerable flexibility and scalability of far memory access. 

\textbf{DM Architectrue:} Disaggregated memory architecture refers to the organization ways of Computing Nodes (CM) and Memory nodes (MN). There are many ways for computing nodes to access external memory spaces in the existing server architecture. We summarize the DM architecture into 4 categories according to the end number of each far memory path. The number of compute nodes and memory nodes involved in the far-memory access path determines the design and system implementation of different disaggregated memory architectures.
%\textbf{DM Architecture:} At the server level, prior works mainly concentrate on the interconnection between computing and memory nodes. DM can be composed of different physical memory pools inside memory nodes by connecting different slots, such as the on-chip memory pool, DIMM memory pool, and PCIe memory pool.  From the perspective of compute nodes, there are two ways to access DM, namely vertical (intra-node) and horizontal (i.e., inter-node) far memory access. In addition, DMs can be scheduled in centralized and decentralized ways. 

%\textbf{Cluster infrastructure: } At the cluster level, Composable Disaggregated Infrastructure (CDI) \cite{cdi} maintains disaggregated memory, disaggregated storage, and disaggregated network \cite{sv-sdc} and reorganizes necessary resources. Another widely recognized framework is the software-defined hardware infrastructure (SDHI). A cluster comprising multiple monolithic machines is established in this setup, underpinned by a software-defined environment (SDE) in the Software-Defined Data Center (SDC).

\textbf{FM System/Runtime:} On top of disaggregated memory architectures, runtime design and optimization strategies for far memory access is crucial. They provide the necessary support for high-performance far memory access and efficient memory resource scheduling. We summarize the software environment of disaggregated memory, including system-level OS adaption, runtime design and task scheduling. These systems are designed to support highly flexible resource usage, low overhead data processing, and high task/data throughput.

%Architecture refers to the way the computing nodes and memory nodes are organized. We describe the architecture for the cluster level and the server level. At the cluster level, there are homogeneous and heterogeneous DM architectures. 
% The fundamental parts of the architecture, i.e., the way it is organized, are \textit{ compute nodes} and \textit{memory nodes}. 

%\textbf{Far memory environment design:} Far memory software environment provides resource management and abstraction for user-level applications is implemented in the operating system level and the virtualization level. The key operation is to handle data swaps between local and remote, i.e., swapping the cold data out and the required data in. The coordination of the local memory access (LMA) and remote memory access (RMA) is required. In addition, resource management of computing, memory, network, and storage resources needs to consider the application sensitivity, direct storage access and parallelism design.
%Traditionally, data are organized as pages and handled by a swap mechanism, including a swap front end and a backend. 

% \textbf{Runtime designs: }
% %following existing computing modes, including the in-memory and the out-of-core computing modes???
% %We believe that there will be more far memory systems and runtime designs for today's memory-hungry large language models.

\textbf{Application Optimization:} There is a trend towards adapting and redesigning far memory systems for specific scenarios, i.e., service-aware far-memory design, including serverless, elastic computing, fault-tolerant computing, and power-optimized far memory systems. In addition, related works also design corresponding specialized far-memory systems for dedicated big data tasks, such as graph processing and neural network training and inference. 
Application acceleration mainly adopts data management in the computing node, which considers how to identify, partition, cache, and prefetch the cold data at fine-grained granularity.  Domain-specific far memory system design can optimize the performance of elastic computing, graph processing, and DNN-supported systems.

%The design of the system and the improvement of the task-oriented scenarios require consideration of data relation, the computation mode, and the data communication operations.??? 什么是data relation？

% \textbf{Far memory in heterogeneous environment:} Heterogeneous computing also embraces far memory systems for memory extension. 

% There are two main relationships of the processed data on different memory spaces, especially on a fast and a slow memory/storage space. 
% \begin{itemize}
%     \item Inclusive: This means the system often caches or buffers a small proportion of data from the slow memory entities to the fast memory entities.   
%     \item Data mapping/swapping (Exclusive):
% \end{itemize}

% We systemically conclude the conception of \textit{far memory access }to make it more clear.

 %The performance of Inter-FM and Intra-FM  is getting closer so combining Inter-FM and Intra-FM would provide better performance per bit. 
% \subsection{Paper Overview}

% In this paper, the disaggregated memory system includes the heterogeneous disaggregated memory resource organization at the architecture layer, the adaptation and construction of the far-memory system at the system layer, and the design and optimization of the far-memory access patterns at the application layer. 

%Figure \ref{fig:structure} shows the system design level of disaggregated memory. 

%% file: Contents/3-Architecture.tex
\section{DM Hardware}\label{sec-hardwareusage}
Disaggregated memory requires new architecture/hardware designs for better elasticity and efficiency. This section first introduces cluster-level and server-level DM architectures. We then discuss DM scheduling methods. 

To figure out how to deploy disaggregated memory on physical servers, this section introduces hardware that can be recognized as disaggregated memory, i.e., far memory. General-purpose computing nodes, such as CPUs and GPUs, can access far memory devices directly across hardware bus and slots, or through a network card. Thus, the key components of disaggregated memory are memory devices and network devices, i.e., \textbf{disaggregated memory} = (\textbf{network connections}) + \textbf{memory devices}, The "()" refers that network connections can be not included. 
We also provide a performance comparison of advanced memory, storage, and network devices.

% Server-level DM architecture can be viewed as a basic component of cluster-level DM architecture. 

\subsection{Memory devices}\label{subsec-memorydevice}
%取下边的素材
Memory devices usually communicate with CPUs through on-chip buses and interfaces. The first column of Table \ref{tab:memory-connection-device} shows the commercially used memory devices, the slots, channel, and protocol. 
We compare different memory and memory connection devices according to the plug-in slots, data channels, and data transfer protocols, as shown in the rows of Table \ref{tab:memory-connection-device}. Slots refer to the way memory devices and compute motherboards are interconnected and are usually hardware interfaces. The channel refers to the available data channels decided by the devices or the slots. Based on typical slots and channels, supported data transfer protocols are also required.

Generally, there are two types of memory device that can act as disaggregated memory, with different data persistence capabilities. \textit{Volatile Random Access Memory} (RAM) includes both CPU DRAM, and GPU DRAM. The high-bandwidth memory (HBM)  \cite{hbm} and hybrid memory cube (HMC)  \cite{memory-network} are proposed for higher density of DRAM memory. In addition, \textit{Non-Volatile Random Access Memory} (NVRAM, short as NVM) can store data even when the power is off. Flash memory is a type of NVRAM commonly used in the design of SSDs  \cite{sumsung-ssd}. Persistent Memory (PM) is a type of NVRAM; taking the Intel Optane NVM   \cite{persistentmemory} as an example, PM is located in the system DIMM slots with an extra power control module. Furthermore, academia and industry are researching and experimenting with promising candidates for next-generation memory ("x"RAM), including magnetoresistive memories (MRAM), resistive memories (ReRAM), phase-change  memories (PCRAM), and ferroelectric memories (FeRAM), etc. 

%Domain-specific material storage devices can bring significant performance improvement on memory devices. 

%DRAMs are facing the challenge of shrinking in size, with DRAM now approaching the limits of shrinkage and the transition to 3D architectures underway. 
%In addition, CXL (Compute Express Link)   \cite{cxl} is one of the volatile RAM, a new attractive technology that supports disaggregated memory devices to communicate with CPU and local memory through PCIe interfaces.  

% For example, persistent memory such as NVM (non-volatile memory) uses DIMM slots to communicate with DRAM memory and CPUs. 

% SSD storages use PCIe, M.2, U.2, SATA ports to connect with the main board. 
% The bus-based memory device connection method refers to the direct insertion of memory devices by using high-channel slots or interfaces on the chip. These are usually on the same chip, thus providing the widest data access channels and the lowest data transfer latency for each memory chip. This requires two main components, the hardware slots and the software protocol.

% \begin{figure*}[t]
%      \begin{minipage}{0.6\linewidth}
%         \centering
%         \includegraphics[width=\linewidth]{figure/io-bw-trend.pdf}
%         \caption{I/O bandwidth trend doubles every 3 years   \cite{pcie6.0}. }
%         \label{fig:io-bw-trend}    
%     \end{minipage}
% \end{figure*}

\textbf{DIMM slots and protocols}: The commercially used DRAMs are Double Data Rate (DDR) DRAMs. With the expansion of the bit width of data readable by DDR and the upgrading of hardware structure, DDR2 can accomplish 4-bit data pre-reading, DDR3 is 8-bit, DDR4 is 16-bit, DDR 5 is 32-bit, and DDR6 is 64-bit. Every next version of DDR maintains double data bit width and thus has double data transmission bandwidth. During this evolution process, the working voltage of DDR has also been gradually lowered, and the working frequency has been gradually increased. Adding the block number can also add data bandwidth. With a single block achieving 34 GB/s in DDR4, one can have at least 200GB/s data bandwidth on eight DDR4 banks. The latest commercial DDR version is DDR5 (until 2024), while DDR6 is coming soon\cite{intel-PMC}. 
    
\textbf{PCIe slots and protocols}: PCIe bandwidth is keeping an increasing trend, which is roughly faster than double times every three years  \cite{pcie6.0}. For example, PCIe 3.0 is twice as fast as PCIe 2.0, and PCIe 5.0 is twice as fast as PCIe 4.0 with backward compatibility. Currently, the commonly used duplex-lane PCIe 4.0 x16 version commonly has about 64GB/s of bandwidth. PCIe Gen 5.0 with total throughput of 128GB/s and PCIe 6.0x16 with 256GB/s bandwidth is considered. The rapid development of PCIe has allowed disaggregated memory devices to approach the bandwidth of traditional DRAM. In addition, PCIe also has better interface scalability with extendable roots.
%As Figure \ref{fig:io-bw-trend} shows, 
%This indicates that disaggregated memory devices based on PCIe slots have a brighter future. 
    
\textbf{Dedicated slots and protocols}: Recent works propose direct connection devices, such as NVLink, which can support direct memory communication between GPUs  \cite{nvlink}. Tavallaei et al. propose an open accelerator module (OAM)  \cite{oam} supporting various interconnect topologies for new types of hardware accelerators (such as GPU, FPGA, ASIC, NPU, TPU, IPU, et al). IBM Power 9 allows direct memory access between servers via specific cables and the OpenCAPI protocol   \cite{opencapi}. Recently, Intel has developed CXL CPU motherboards, which are based on the enhanced protocol PCIe 5.0 and are capable of supporting approximate access latency for cross-core (socket) accesses, as well as interconnecting memory pools with accelerator pools  \cite{cxl}.  Several works   \cite{optically-jpdc2022, opticalFiber} have proposed the use of electronic signals instead of photonic signals for memory interconnections. 

%Although the current dedicated slots may be too expensive to deploy, they provide a new way to break the traditional architecture limits, adding more "highway" data transfer paths.

\begin{table}[t] \small
\centering % 使表格居中
\caption{Memory and Connections that can be organized as Disaggregated Memory.} % 表格标题+7\label{tab:my_table} % 用于引用表格的标签
\vspace{-0.2cm}
\label{tab:memory-connection-device}
\begin{tabular}{|c|c|c|c|c|}
\hline
\textbf{Components} & \textbf{Device }& \textbf{Slot} & \textbf{Channel} & \textbf{Protocol} \\ \hline
\multirow{8}{*}{\begin{tabular}[c]{@{}c@{}} \textbf{Memory}\\ \textbf{Device} \\ \textbf{(Sec. \ref{subsec-memorydevice})} \end{tabular}} 
  & DRAM  &  DIMM/HBM/HMC  &  Bus  &  DDRx, (x is up to 6)  \\\cline{2-5} 
  & GPU DRAM & PCIe and NVLink  & NVLink and NVLink switch & GDDR5, GDDR6 \\\cline{2-5} 
  & xPU DRAM & OAM  & OAM on x16 links & PCle Gen5/6 x16\\\cline{2-5} 
  & NVM & \makecell[c]{DIMM with \\ extra power support}  &  Bus & Customized \\\cline{2-5} 
  & Persistent memory & DIMM or PCIe  &  Bus or PCIe switch & Customized \\\cline{2-5} 
  & CXL memory & PCIe/DIMM &  SATA, PCIe x32, x64, etc., bus & CXL 1.0, 2.0, 3.0, etc. \\\cline{2-5} 
  & SSD & PCIe, M.2, U.2 &  SATA, PCIe x4, x8, etc. & AHCI, NVMe \\\cline{2-5} 
  &  \makecell[c]{"x"RAM \\ (MRAM, ReRAM, \\ PCRAM, FeRAM, etc.)}  & DIMM or PCIe  & Bus or PCIe switch & Customized \\\hline

\multirow{4}{*}{\begin{tabular}[c]{@{}c@{}} \textbf{Network}\\\textbf{Connection}\\  \textbf{Device} \\ \textbf{(Sec. \ref{subsec-memoryconnection})}\end{tabular}} 
& NIC & PCIe & PCIe x4, x8, x16, etc. & TCP/IP \\\cline{2-5} 
& RDMA NIC & PCIe  & PCIe x4, x8, x16, x32, etc. & RDMA\\\cline{2-5} 
& DPU (for SmartNIC) & PCIe & PCIe x16, x32,  etc. & Customized\\\cline{2-5} 
& FPGA (for SmartNIC) & PCIe & PCIe x16, x32,  etc. & Customized\\\cline{2-5} 
\hline
\end{tabular}
\vspace{-0.1cm}
\end{table}

%\textbf{Cache-coherent CXL memory:} 

%\textbf{Dedicated memory networks:} 

\subsection{Network connections}\label{subsec-memoryconnection}
The network connections used for disaggregated memory (or far memory) mainly involve network interface cards (NIC) and programmable-chip-based smart NICs. Details are shown in Table \ref{tab:memory-connection-device}. 

 % Figure \ref{fig:bwandlatency} gives the detailed difference of memory bandwidth and read/write latency of FM backends. 

\textbf{RDMA NIC}: 
RDMA technology is one of the main methods of far memory access  \cite{fastswap,xmempod,infiniswap,aifm,canvas-nsdi23}. RDMA is widely used in data-intensive and compute-intensive scenarios as a remote memory direct access technique that bypasses the kernel. RDMA has three network protocols: Infiniband, RoCE, and iWARP. The main manufacturer of InfiniBand network is Mellanox, which is now acquired by NVIDIA. RoCE (RDMA over Converged Ethernet) is a network protocol that allows RDMA to be performed over Ethernet with lossless Ethernet for low-latency operation and complex configuration process. The AIPs provided by OFED  \cite{ofed} and IB face problems with usage, performance, and security. Thus, some works provide high-level programming interfaces   \cite{farm-nsdi14,remote-regions,lite}.

%RDMA is characterized by zero-copy, stack offloading. RDMA uses IB verbs to read and write data by establishing information transmission channels. 

\textbf{Smart NICs}: By designing high-speed on-chip memory networks with routing logic, one can deploy thousands of memory modules on a single chip and realize on-chip interconnections  \cite{stringfigure}. Meanwhile, there have been works to design customized smart NIC devices to replace generic NIC devices and reduce the overhead of far memory access  \cite{clio}. Also, there is some work designing programmable switches  \cite{mind} to work with server-level RDMA or smart NIC devices for cabinet-level memory interconnections. MIND   \cite{mind} places memory management logic in the network fabric, and they find that centralizing memory management in the network permits bandwidth and latency-efficient realization of in-network cache coherence protocols. At the same time, programmable switch ASICs support other memory management logic at line rate.

\subsection{Characteristic of DM Hardware}\label{subsec-performance-analysis}
\begin{figure*}[t]
     \begin{minipage}{\linewidth}
        \centering
        \includegraphics[width=\linewidth]{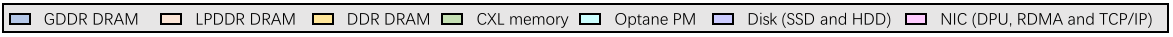}
        \includegraphics[width=\linewidth]{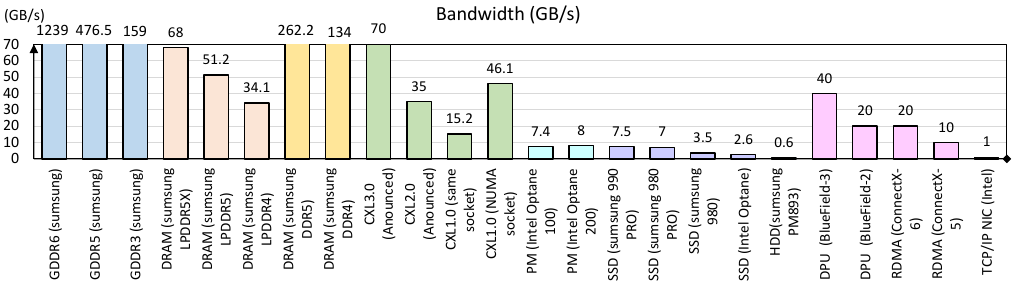}
        \caption{Memory bandwidth of different far memory devices, including memory, storage and network card  \cite{sumsung-ssd,nvidia-dpu,nvidia-rdma}.}.
        \label{fig:bandwidth}    
            \vspace{-0.3cm}
    \end{minipage}
     \begin{minipage}{\linewidth}
        \centering
        \includegraphics[width=\linewidth]{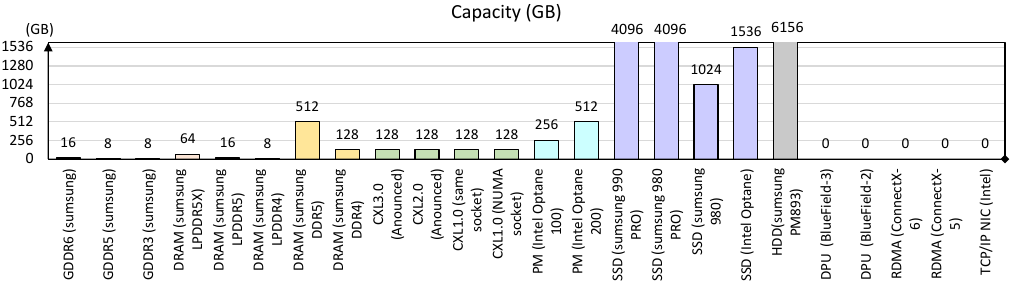}
	\caption{Memory capacity of different far memory devices, including memory, storage and network card.}
	\label{fig:capacity}

    \end{minipage}
         \begin{minipage}{\linewidth}
        \centering
        \includegraphics[width=\linewidth]{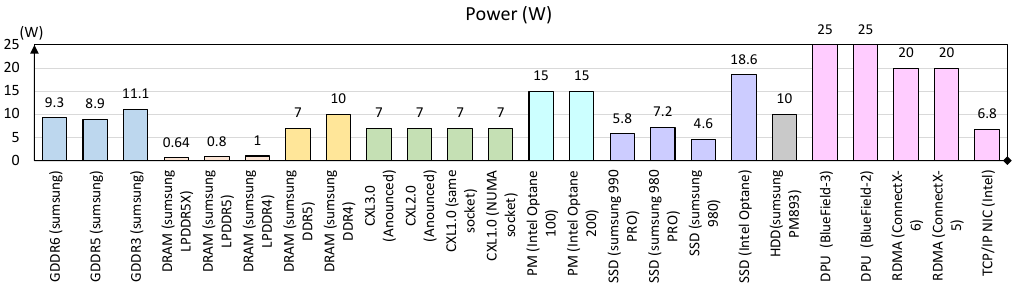}
	\caption{Power usage of different far memory devices, including memory, storage and network card.}
	\label{fig:power}
 \vspace{-0.3cm}
    \end{minipage}
    \vspace{-0.3cm}
\end{figure*}

%一段话分析一个数据图，每个数据图讲 1）不同的内存设备数据差异2）同一种设备一直推出新的型号，性能有差别
We illustrate the performance of the mentioned devices with different versions in bandwidth (Figure \ref{fig:bandwidth}), capacity (Figure \ref{fig:capacity}), and power (Figure \ref{fig:power}). In these figures, GDDR is the main memory for GPUs. LPDDR DRAM is used in edge devices such as mobile phones and DDR DRAM serves as the main memory for servers. CXL memory, persistent memory, and SSDs are memory devices that can store data. DPUs, RDMA, and NICs represent network cards used for data transfer with optional data caches or control logic.

\textbf{Bandwidth analysis: }
Available data bandwidth is one of the key metrics of data access performance.  
As local memory, DRAMs have large memory bandwidth with hundreds of GB per second (GB/s) to support high-performance data access for computing units. GDDR memory, designed for GPUs, delivers the highest memory bandwidth to handle the large-scale parallel processing demands. LPDDR DRAMs for mobile devices with relatively lower power consumption and less data bandwidth. Persistent memory and SSDs are non-volatile storage devices with large memory capacities but offer lower bandwidth. Oftentimes, new-generation memory device exhibit wider bandwidth. For example, CXL 3.0's 70GB/s far exceeds CXL 1.0's 15.2GB/s in the same socket physical interface. An important trend is that networking devices and storage devices are evolving at a rapid pace, much faster than native DRAM-based DDR. while individual DRAMs still lead in performance today, their overall bandwidth may reach the upper bound due to limited on-chip DIMM slots. State-of-the-art interconnects is establishing scalable plug-in interconnect methods based on PCIe, OCP Accelerator Module (OAM), and others. 
% CXL, RDMA, and DPU are designed to enhance data transfer efficiency through network-based interconnects, adopted as high-performance disaggregated memory device in recent research works. They have lower bandwidth than local memory but much higher bandwidth compared with storage-based far memory.
% Persistent memory and SSDs, non-volatile storage types, have lower bandwidth compared to other memory types, as they are primarily designed for data maintenance. Newer memory devices generally exhibit wider bandwidth, leading to improved performance. For example, CXL 3.0 offers 70GB/s, far exceeding CXL 1.0's 15.2GB/s in the same physical interface.

\textbf{Capacity analysis:}
GDDR memories used in GPUs have higher bandwidth than DDR DRAMs used in servers, but with lower capacity due to specific application requirements. GDDR memory and LPDDR DRAM can work with lower capacities compared to the DDR DRAM in servers, since GPUs and mobile devices focus on performance and power efficiency, whereas servers need larger memory capacities to handle high data processing demands. CXL is an interconnect technology that not only enables data transfers but also incorporates a fixed memory capacity of 128GB. Non-volatile storage types, such as persistent memory and SSDs, are known for their large storage capacity compared to other memory devices. 

%Main memory processes a smaller storage capacity in comparison to non-volatile memory types. While DRAM shows lower capacity than Solid-State Drives (SSDs), its bandwidth is much higher. 

%Interconnect technologies are designed solely for data transfer between components. They do not have an inherent memory capacity limitation. 

% Except for CXL, which remains consistent across models, all other device categories have increased in capacity with updates.

% Reviewing the top capacities for different devices, HDD(samsung PM893) offers with considerable capacity of 6156 GB. Samsung's 990 Pro and 980 Pro SSDs both deliver 4096 GB. Other devices such as DRAM, PM, CXL, and GDDR have lower capacities at or below 512 GB. RDMA, TCP/IP NICs, and DPUs are not equipped with memory capacity as their functions are for data processing and networking.
% 
\textbf{Power analysis: }
Power consumption varies significantly across different types of memory and interconnect technologies, reflecting their different design goals. For instance, DPU has a power draw of around 25 W. However, a LPDDR DRAM consumes less than 1 W.
When assessing the power consumption of a system, we need to consider the total power consumption rather than focusing solely on individual device power. For instance, while a single LPDDR DRAM may consume less power, there may be numerous such modules in a mobile device, leading to a significant cumulative power.

\section{DM Architecture}\label{sec-architecture}

Disaggregated memory architecture refers to the organization ways of Computing Nodes (CM) and Memory nodes (MN). There are many ways for computing nodes to access external memory spaces in the existing server architecture. Generally, it is commonly used that far memory path =  processing units + local memory + connections + far memory. 

\begin{figure*}[t]
    \centering
    \includegraphics[width=0.9\linewidth]{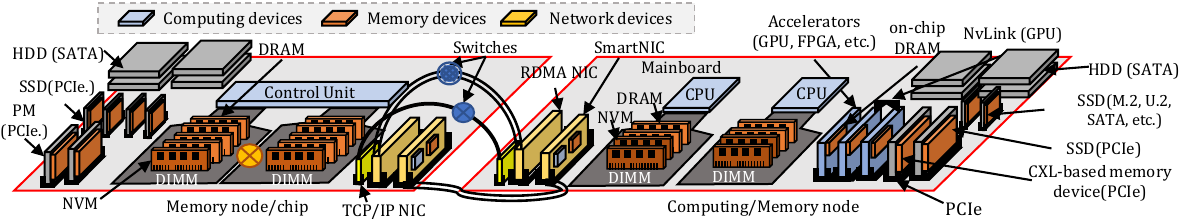}
    \caption{Physical disaggregated memory testbed with various memory and network devices. }
    \label{fig:memorypool}
    \vspace{-0.3cm}
\end{figure*}

In real physical machines, FMs can be on the same machine or on different machines interconnected by a network.  Figure \ref{fig:memorypool} shows an example of the hardware deployments of a computing node and a memory node. DM can be built on a single server or paired servers. 
Compute nodes usually include computation units, caches, and local memory, while memory nodes mainly include memory control units, memory interconnects, and far memory. Far memory space often refers to isolated free-accessed memory regions located on certain memory devices or servers.

In this section, we summarize the DM architecture into 4 categories according to the end number of each far memory path. The number of compute nodes and memory nodes involved in the far-memory access path determines the design and system implementation of different disaggregated memory architectures. Similarly to the mapping of instruction and data stream in CPU architecture, we use single/multiple computing node (CN) Single/Multiple memory node (MN) to describe the disaggregated memory architecture.

% DM is secondary/extended memory components that are relatively "far" from the local computing resource and local memory.

% \subsubsection{Far memory access ways} The way to access a far memory in the disaggregated memory pool matters. DM is often described as secondary/extended memory components that are relatively "far" from the local computing resource and local memory.

% Far memory access paths can be implemented in a variety of ways. Existing work has been able to "assemble" DM architectures for various scenarios by combining commercial or customized memory devices and memory connections. These various far memory access paths also require efficient management methods to fully utilize the corresponding resources. 

%We introduce memory devices and memory connection methods. We also classify far memory systems based on device and connection types.

% , as shown in  Figure \ref{designlevel-fma}-(2).

\subsection{Single CN Single MN} 
The basic DM architecture is one single computing node with one single memory node.  The computing node can be CPU processors, GPU processors, or any other accelerators. We use FM paths to represent the data paths from one computing node and a memory node.   

\subsubsection{CPU far memory}

Currently, there are mainly three types of far memory access paths: the memory-based FM paths, network-attached FM paths, and controller-involved FM paths, as shown in Table \ref{tab:3-dm-type}.

%Existing efforts on designing memory connection architectures with available far memory paths include the memory-based disaggregated memory (DM), network-attached DM, and memory controller-involved DM, as shown in Table \ref{tab:3-dm-type}. 

 \textbf{Memory-based FM path:} This type of FM often utilizes persistent memory like NVM and fast storage like SSD, as shown in the second column of Table \ref{tab:3-dm-type}. These works extend the memory inside each server, which are called \textit{vertical} or \textit{intra-node} far memory. Some works focus on scaling one's own memory to special memory nodes with a high-capacity NVM in the rack   \cite{pDPM-persistentmemory,persistentmemory,anderson2020assise,zhang2022ford,lee2022dinomo}. They try to design persistent, replicated, fault-tolerant coherence protocols on distributed file systems for Optent DC persistent memory modules. Many works adopts SSD  \cite{asplos19software,scm-atc,tmo,vswapper,hybridswap,xmempod}
    as data offloading space. These works take the fast data read of SSD to accelerate I/O operation and design related resource management methods for high efficiency. 

%Random Access Memory (RAM) often acts as the far memory space and plugs in the memory slot directly. 

\textbf{Network-attached FM path:} It often connects memory devices through network cards, as shown in the third column of Table \ref{tab:3-dm-type}. These works extend the memory between servers, which are called \textit{horizontal} or \textit{inter-node} far memory.  Many works utilize RDMA NIC  \cite{fastswap,icdcs19,infiniswap,fargraph,rethinking-asplos21,zhang2022ford,aifm,farm-nsdi14,lite,remote-regions} as far memory. They explore the advantages of the RDMA protocol, including the non-CPU-involved computing workflows, one-sided operations, and variable data transfer chunk size. Some recent works adopt smart NICs implemented by FPGA or DPU as far as memory   \cite{clio,Thymesisflow}. These works offload some data processing workloads to the low-power devices to improve memory efficiency. However, the proposed smartNIC-based works often suffer performance bottlenecks compared with RDMA   \cite{bluefield}. For cross-rack far memory access, a programmable switch is required. For example, one can place memory management logic in the network fabric and utilize a PCIe switch to enable rack-level high-bandwidth memory communication  \cite{mind}. 

% The fabric-attached far memory usually communicates with CPUs and local memory by inter-connected fabrics and networks. They use the message-passing mode that get data from the network. For instance, classic distributed computing system use RPC model on TCP/IP network cards to fetch data on the remote memory  \cite{grpc}. Today's far memory system usually uses the high-speed direct memory access(DMA) model on RDMA(Remote Direct Memory Access) network card through copper cables with Infiniband switches to access remote memory. 

\textbf{Controller-involved FM path: } DM devices based on reconfigurable hardware involve specific controllers for highly efficient memory management, , as shown in the forth column of Table \ref{tab:3-dm-type}. For example, String Figure   \cite{stringfigure} builds a large memory pool with thousands of memory nodes and tens of CPUs, designing a hybrid routing protocol and a set of network reconfiguration mechanisms. Beacon \cite{beacon-micro22} proposes a memory management framework to enable memory expansion with unmodified CXL-DIMMs. Mind   \cite{mind} uses a control plane to track the total amount of memory allocated on each memory blade with leveraging its global view of allocations for load-balanced memory pool management.

\begin{table}[t]
\centering 
\caption{Taxonomy of far memory (FM) access paths.}
\label{tab:3-dm-type}
    \vspace{-0.3cm}
 \begin{tabular}{|c|c|c|c|}
%\begin{tabular}{|m{1cm}|m{3cm}|m{5cm}|m{5cm}|}
\hline
\textbf{Type} & Memory-based DM & Network-attached DM & Controller-involved DM \\
\hline
\raisebox{1cm}{\textbf{Arch.}} & 
\includegraphics[height=2.2cm]{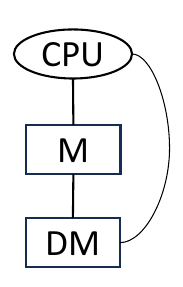} & \includegraphics[height=2.2cm]{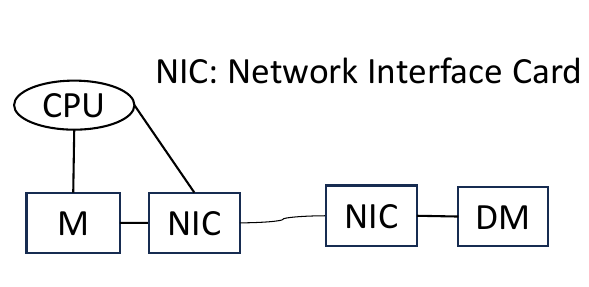} & \includegraphics[height=2cm]{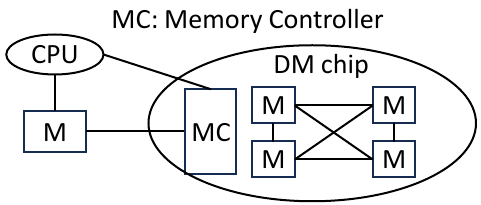} \\
\hline
\textbf{\makecell[l]{Ref. \\ (CPU)}} 
&\makecell[l]{Persistent memory:  \\   \cite{pDPM-persistentmemory,persistentmemory,anderson2020assise,zhang2022ford,lee2022dinomo} \\ CXL-based memory: \\ \cite{CXLbasedSSD,pond-cxl} \\ Disk and SSD: \\   \cite{asplos19software,scm-atc,tmo,hybridswap,xmempod} } 
& \makecell[l]{RDMA and DRAM: \\   \cite{fastswap,icdcs19,infiniswap,fargraph,aifm,farm-nsdi14,lite,remote-regions} \\
RDMA and NVM:  \cite{zhang2022ford} \\
FPGA and DRAM:   \cite{clio,Thymesisflow}}
&  \makecell[l]{On-chip memory network:  \\  \cite{stringfigure,disaggregated-mem-isca09,zombieland,Thymesisflow,griffin} \\ \cite{mind,rethinking-asplos21,coarse-hpca22,clio} }\\
\hline

\textbf{\makecell[l]{Ref. \\ (GPU)}} 
&\makecell[l]{CPU memory:  \cite{scm-uvm,wang-grus,dragon-sc18} } 
& \makecell[l]{NVLink-based memory \cite{nvlink} \\RDMA: \cite{nvidia-rdma} }
&  \makecell[l]{SSD: \cite{bam} } \\
\hline
\end{tabular}
    \vspace{-0.3cm}
\end{table}

\subsubsection{GPU far memory}
Heterogeneous far memory systems and runtime designs are new trends for today's memory-hungry workloads, such as training and inference of large foundation models. These workloads often rely on general-purpose accelerators, i.e. GPUs for high parallelism. Therefore, GPUs require orders-of-magnitude higher memory bandwidth than traditional CPU-based systems \cite{2022ai-hpca}. However, the memory space of GPU is still limited and the capacity of such high-bandwidth memory tends to be relatively small. There are several GPU far memory access paths, including CPU memory, NVLink-based remote GPU memory, direct-storage-based storage space, direct-rdma-based remote memory, etc. Thus, GPUs can access far memory through several paths, including unified memory (UM), GPU direct storage (GDS), and GPU Direct RDMA. 

\textbf{Unified memory:} In GPU far memory architecture, Unified Virtual Memory (UVM) is an essential way to integrate GPU and CPU memory spaces. This arrangement facilitates data sharing between CPU and GPU, with no transfer exposed to apps and users. Recent work focuses on leveraging CPU memory for specific purposes and optimizing current management strategies.
Swap Advisor \cite{swapadvisor} proposes a general swapping mechanism that can swap deeper and wider neural networks on limited GPU memory. The memory consumption is reduced by smartly swapping the model parameters that are updated at the end of each iteration and used by the next iteration. To optimize GPU's memory access overhead, Grus \cite{wang-grus} co-utilizes different memory-sharing operations of CUDA UVM when running graph workloads. Shao et al.  \cite{shao-icpe} investigated the implications of GPU UVM oversubscription. They evaluate the performance sensitivity of applications with over-subscribed memory size on GPU. Some works adopt memory efficient techniques with static and dynamic scheduling strategies to execute large-scale graph random walk on GPU and CPU memory\cite{wang-skywalker,flowwalker-mei}.  Soroush et al  \cite{co-optimizing} explored the hidden performance implication of GPU memory management methods and developed a performance model that can predict system overhead for each memory management method.  

\textbf{GPU direct memory access:}
The integration of GPU memory with high-speed storage devices like SSDs and NVMe drives is another frontier in GPU far memory management. Besides storage devices with sufficient bandwidth for memory management, novel network technology and hardware like RDMA allow data access faster and lower-costly. Current work focuses on reducing data transfer volume by achieving higher offload and prefetch efficiency and achieving higher bandwidth and lower latency with hardware and software methods.

GPU Direct-storage (GDS) techniques \cite{direct-storage} create a direct data path between local or remote storage, such as NVMe or NVMe over Fabrics (NVMe-oF), and GPU memory. By enabling direct-memory access (DMA) near the network adapter or storage, it moves data into or out of GPU memory—without burdening the CPU. 
GPUDirect RDMA  \cite{direct-rdma} is a technology introduced in Kepler-class GPUs and CUDA 5.0 that enables a direct path for data exchange between the GPU and a third-party peer device using standard features of PCI Express.  BAM  \cite{bam} proposes a big accelerator memory system that enables GPUs to orchestrate high-throughput, fine-grain SSD storage access. It mitigates I/O traffic amplification by enabling the GPU threads to read or write small amounts of data on-demand through highly concurrent I/O queues.

\subsection{Single CN Multiple MN} 

One can utilize multiple far memory nodes for a single computing node to improve data throughput.The compute nodes need to be carefully designed to utilize external heterogeneous memory resources. If there are multiple far memory devices, one may hierarchically utilize the hybrid memory devices or use the multiple far memory backends in parallel.  

\subsubsection{Hierarchical hybrid memory}
Hybrid memory combining fast and slow heterogeneous memories provides a promising direction to increase memory capacity. 

\textbf{General-purpose hybrid memory design: }Some works focus on hybrid memory management for general works.  Hybrid${^{2}}$ \cite{hybrid2} considers a hybrid memory system composed of memory technologies with different characteristics that use only a small fraction of the near memory as a DRAM cache and further leverages DRAM cache for large data migration. Hybrid TLB Coalescing \cite{park2017hybrid} proposes a novel HW-SW hybrid translation architecture, which can adapt to different memory mappings efficiently. UH-MEM \cite{li2017utility} proposes utility-based hybrid memory management, a new page management mechanism for various hybrid memories that systematically estimates the utility of migrating a page between different memory types and uses it to guide data placement. 

\textbf{Persistent-memory-based design: }Some works focus on hybrid DRAM and NVM  memory hierarchy. 
Memos \cite{liu2019hierarchical} proposes a memory management framework that can hierarchically schedule memory resources over the entire memory hierarchy, including cache, channels, and main memory comprising DRAM and NVM simultaneously. 
Panthera \cite{wang2019panthera} proposes a semantics-aware automated memory management technique for Big Data processing over hybrid memories (DRAM and NVM) by analyzing user programs to infer their coarse-grained access patterns.MNEME \cite{song2020exploiting} proposes an architectural change where each long bit line in DRAM and NVM is split into two segments by an isolation transistor. 

\subsubsection{Multi-path far memory }
The coexistence of multiple far memory backends is a boost for offloading efficiency and system flexibility. 
Multiple far memory access paths on multiple memory backends (multi-end DM) may become the future solution for improving data throughput. 
%However, the induced complexity needs to be carefully handled. 

\textbf{Multiple far memory paths for CPU:} 
XMemPod \cite{xmempod} established access to the memory of other VMs through host- and RDMA-coordinated memory hierarchy. On the other hand, TMO \cite{tmo} implements its memory pool through zswap and swap. Buddy Compression \cite{buddycompression} compresses and allocates data between GPU device memory and buddy memory such as NVLink \cite{nvlink}, to reduce reallocations within device memory and adapt to compressibility changes.
XDM\cite{wang-xdm-sc24} design parallel multiple far memory backends and can adaptively switch the backends with dedicated far memory path configurations.

\textbf{Multiple far  memory paths for GPU:} 
Buddy Compression \cite{buddycompression}  leverages compression to utilize a larger buddy memory from the host or DM, effectively increasing the memory capacity of the GPU. It compresses the offloaded memory locally as compressed memory and utilizes buddy storage with alternatives, including CPU memory, DM, unused peer GPU memory, and NVMe SSDs. ZeRO-Infinity  \cite{Zero-infinity} is a novel heterogeneous system technology that leverages GPU, CPU, and NVMe memory to allow for unprecedented model scale on limited resources without requiring model code refactoring. G10 \cite{g10-micro23} scale the GPU memory capacity with transparent data migrations by integrating the host memory, GPU memory, and flash memory into a unified memory space.  GMT\cite{gmt-asplos24} design tiered GPU Memory systems with 3-tier hierarchy comprising GPU memory, host memory and SSDs and selectively chooses the provided far memory paths.

\subsection{Multiple CN Single MN} 

% \begin{figure}[t]
%     \centering
%     \includegraphics[width=\linewidth]{figure/designlevel-fma-v2.pdf}
%     \caption{The design scope of far memory system and far memory access on disaggregated memory pool. }
%     \label{designlevel-fma}
% \end{figure}

\begin{figure}[t]
    \centering
    \includegraphics[width=0.7\linewidth]{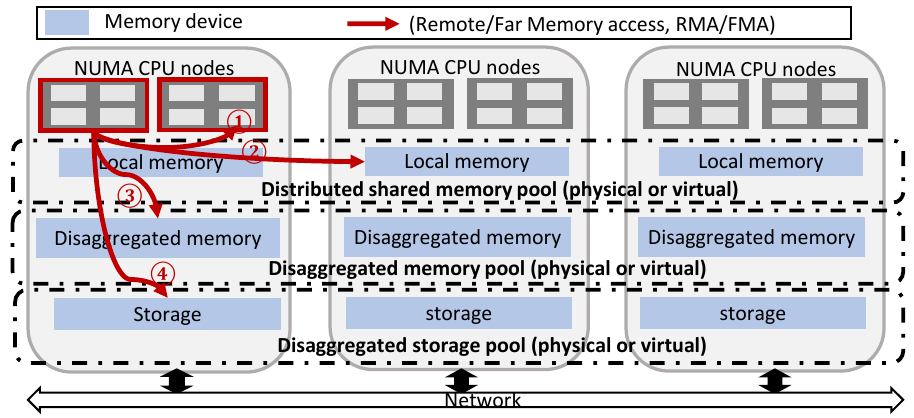}
    \caption{Typical far memory access paths on disaggregated memory. }
    \label{fig:virtualized-pool}
\end{figure}

Oftentimes, one can build a large virtualized memory pool that serves multiple computing nodes. In this way, the computing node may concentrate on the computing and memory access procedure design, which offloads the coordination of the multiple memory nodes to the single memory node, i.e. memory pool. We list four categories of remote memory access (RMA) and far memory access (FMA) methods, as shown in Figure \ref{fig:virtualized-pool}. 

% \subsubsection{Physical memory pool}
% On-chip memory pool:    \cite{stringfigure} 
% NIC-based memory pool:   \cite{beacon-micro22,mind}

% \subsubsection{Virtualized memory pool}
% We classify remote memory access (RMA) and far memory access (FMA) methods into four categories. 
% \textbf{Local and far memory: } DM is often described as secondary/extended memory components that are relatively "far" from the local computing resource and local memory. Local memory is generally the memory space closest to the compute resource. It provides data input and output for compute resources with a very small amount of cache space (usually at the MB level and typically using 64B as the unit of processing data). Differently, \textit{far memory} has more memory space and therefore will provide data input to local memory. Accessing far memory incurs longer latency and lower bandwidth compared with local memory. To improve performance, one may use part of the local memory space for data buffering and it requires coordination with the local memory. 

%Performance degradation of disaggregated memory architecture is always the far memory access from the computing nodes to the memory nodes.
\subsubsection{Far memory pool for CPU}

\textbf{\textcircled{1} Cross-socket RMA/FMA: }  One can access the remote memory space of another CPU sockets through the corresponding Integrated Memory Controller (IMC) on Non-Uniform Memory Access (NUMA) architecture, involving the latency and bandwidth difference between  Local Memory Access (LMA) and Remote Memory Access (RMA)\cite{cpu-less-numa}.
%As the number of CPU cores increases, the Uniform Memory Access (UMA) architecture with shared memory poses a challenge of bus bandwidth pressure and data access conflicts. Non-Uniform Memory Access (NUMA) architecture is proposed to handle different memory devices and CPU cores subordinate to different nodes (or sockets), each of which has its own Integrated Memory Controller (IMC). The IMC bus is used for in-node cross-core memory communication, while the cross-node communication is done through QPI (Quick Path Interconnect). Since the latency of QPI is higher than that of IMC Bus, there is a difference between remote/local memory access by CPU. This is where the concepts of Local Memory Access (LMA) and Remote Memory Access (RMA) come from \cite{cpu-less-numa}.
\textbf{\textcircled{2} Network-based RMA/FMA: } Network-based RMA is usually used to transmit instructions and data through networks, such as TCP or UDP transmission \cite{grpc}. The problem is that data transfer involves many CPU operations, interrupting the original computing on CPUs. Direct Memory Access (DMA) and Remote Direct Memory Access (RDMA) are often adopted for few-CPU-involved features and high-speed data transfers  \cite{rdmarpc-sigcomm}. RDMA-based far memory has an order of magnitude difference in memory access latency and data access bandwidth compared to local DRAM memory \cite{hyfarm-iccd}. Typically, the network interface cards (NIC) of the memory device need to be interconnected with the NIC of the memory device in another server through specific network cables. 
% This may require a special power supply on the motherboard. NIC far-memory access is typically invoked through an application layer interface and is not directly managed by the operating system, thus requiring the application to design better data communication and processing to improve far-memory access performance.
\textbf{\textcircled{3} Memory-based FMA:}
%One can utilize persistent memory or  Compute Express Link (CXL) based memory as DM pool.
Persistent memory devices usually interact directly with the CPU and have their own data processing patterns and programming models\cite{persistentmemory}. 
%Existing work focuses on designing data buffering and mapping to capitalize on memory devices' persistence characteristics and power consumption advantages. 
The newly proposed CXL \cite{cxl} technology is based on PCIe interfaces and it treats memory device as a special NUMA node (memory node) without CPU. %The Linux system has a default NUMA-aware scheme that prioritizes the allocation of local and far memory. When targeting large-memory applications, 
NUMA awareness significantly impacts performance and requires more thread scheduling and memory allocation optimizations \cite{numa-graph}.
\textbf{\textcircled{4} Storage-based FMA:} To ease the performance gap between CPUs and I/O devices, operating systems reserve buffers in memory managed with virtual file systems (VFS) and reduce actual hard disk I/O reads and writes \cite{gridgraph}. 
%When memory resources are insufficient, VFS will automatically evict the buffer. Minor page faults mean that the required pages can be accessed in the buffer, while major faults are not.
Caching and evicting of I/O data can be controlled by LRU list configuration and other system parameters \cite{MGLRU-removefrontswap}. One usually accesses far memory by page swapping, swapping infrequently-used page data to a storage device such as a disk. As storage access performance improves, one can also access far memory directly through an I/O interface \cite{zswapkernel}.

\subsubsection{Far memory pool for GPU}
 Multi-GPU configurations amplify the computational power of many GPUs for serving large-scale applications like Large Language Model (LLM). However, they introduce challenges in managing memory throughout GPUs and accelerate computation and communication. Recent work aims to use a multi-GPU system for distributed deep learning and optimize memory management and communication mechanisms.
COARSE  \cite{coarse-hpca22} is a disaggregated memory extension for distributed deep learning (DL) training on multi-GPU. It proposes dynamic tensor routing and partitioning to fully utilize the non-uniform serial bus bandwidth varied across different cloud computing systems. %It designs a deadlock avoidance and dual synchronization to ensure high-performance parameter synchronization. 
MC-DLA  \cite{micro2018beyond} proposes a memory-centric deep learning system architecture that aggregates a pool of capacity-optimized memory modules within the device-side interconnect for transparent memory capacity expansion. %This architecture allows the accelerators to access the memory nodes via the high-bandwidth links, thus the performance overhead of virtualizing memory can be substantially reduced.
Griffin  \cite{griffin} is a holistic hardware-software solution to improve the performance of NUMA multi-GPU systems. Griffin introduces programmer transparent modifications to both the IOMMU and GPU architecture, supporting efficient runtime page migration.

\subsection{Multiple CN Multiple MN}\label{subsec-cluster-level}
The Multiple CN Multiple MN architecture means that we organize multiple computing nodes with multiple memory nodes on multiple servers. We summarize them into two types, the homogeneous CDM architecture and heterogeneous CDM architecture, as shown in Table \ref{tab:cluster-architecture}. 

% In this subsection, we abstract the clusters from the view of disaggregation.
% 下面这段话反复再讲一些没有什么深度的东西。避免用看似深刻的语言讲没有深度的东西。
%Typically, clusters consist of computing components and memory components. The compute components consist of compute units, hierarchical caches, and local memory, etc. The memory components include the memory control module, read/write cache, storage unit, etc. As a complement of memory components, disaggregated/far memory components can provide additional memory space for high-speed access to compute components. The responsibility of each compute/memory node is to manage the data processing on hierarchical memory (local memory or far memory) of monolithic servers. The responsibility of an architecture is to manage the network (network for computation, network for data) of all the servers in the cluster.

\begin{table}[h] \small
\centering % 使表格居中
% \begin{tabular}{|c|c|c|}
\caption{The two architectures of cluster-level disaggregated memory. In this Table, each Computing components represent the computing units(CPU, GPU, etc.), caches (L1, L2, last-level cache, etc.), and local memory (DRAM, HBM, etc.). Disaggregated memory (DM) represents far/disaggregated devices.} % 表格标题
\begin{tabular}{|c|c|c|}
% \begin{tabular}{|m{0.5cm}|m{6cm}|m{7cm}|}
\hline
\textbf{Type} & \textbf{Homogeneous CDM Architecture} & \textbf{Heterogeneous CDM Architecture} \\
\hline
\rotatebox{90}{\textbf{Cluster Architecture}}& 
\includegraphics[height=3.5cm]{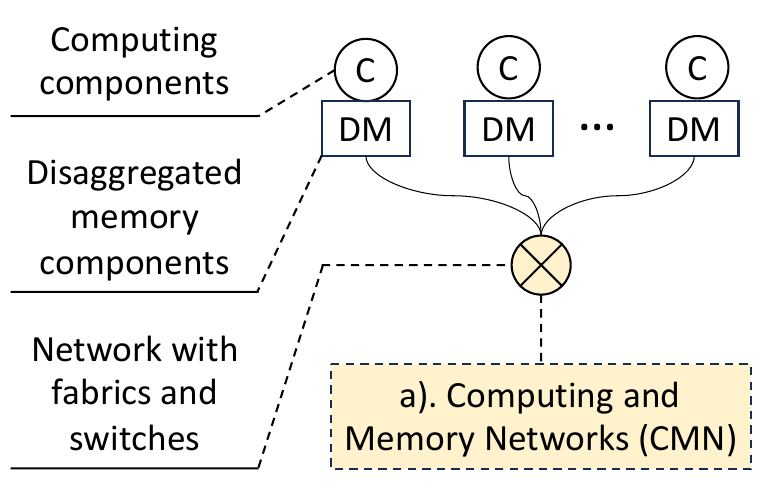} & \includegraphics[height=3.5cm]{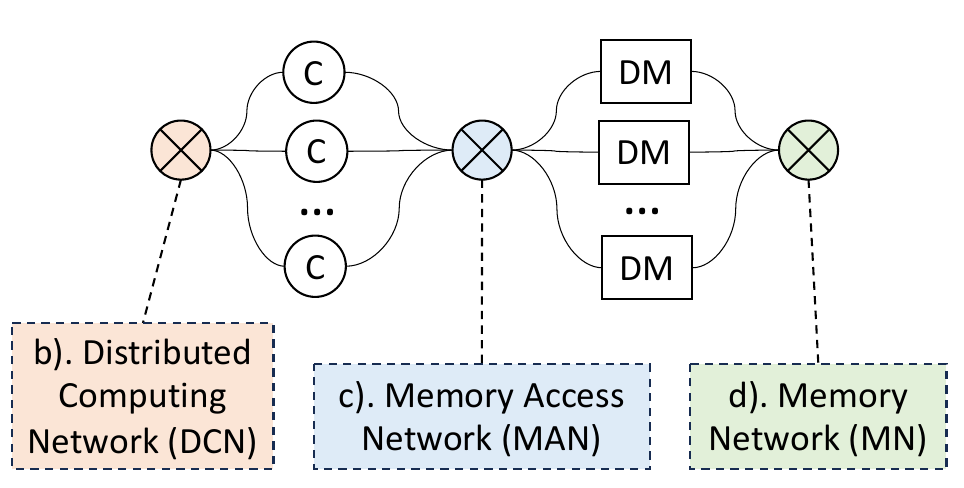} \\
\hline
\textbf{Ref.} &\makecell[l]{ Zswap  \cite{asplos19software} (disk), pDPM  \cite{pDPM-persistentmemory} (NVM), \\TMO  \cite{tmo} (SSD), HyFarM  \cite{hyfarm-iccd} (RDMA and SSD), \\ Xmempod  \cite{xmempod} (RDMA and SSD), \\Pond  \cite{pond-cxl} (CXL), ANNS  \cite{CXL-ANNS} (CXL) }& \makecell[l]{ gRPC  \cite{grpc} (DCN), Tensorflow-RDMA   \cite{tensorflow-RDMA} (DCN), \\ Fastswap  \cite{fastswap} (MAN),  Memliner  \cite{memliner} (MAN), \\ Thymesisflow  \cite{Thymesisflow} (MAN), Memory blade  \cite{disaggregated-mem-isca09}(MAN), \\ String figure  \cite{stringfigure} (MN), Hybrid2  \cite{hybrid2} (MN)} \\
\hline
\end{tabular}

\label{tab:cluster-architecture} % 用于引用表格的标签
\end{table}

\subsubsection{Homogeneous CDM architecture}
%同构架构中，通常是内存节点部署在计算节点上，为计算节点增加额外内存空间，并通过网络将这些内存资源互联。
%在这里我们认为计算部件包括：计算单元、 层级cache、本地内存等；分离式内存部件包括：内存控制模块、读写cache、存储单元等。分离式内存部件能够为计算部件提供额外的高速访问的内存空间。

%则1）每个计算/内存节点主要负责本机器的两级内存（近内存、远内存）的管理，同时2）整个架构需要对网络进行管理（为计算服务的网络，为数据服务的网络）。

In a \textit{homogeneous} CDM, the disaggregated memory resources are usually deployed on the same physical machine as the compute components, which can be interpreted as the compute nodes adding extra memory space. It can achieve high performance, but it requires careful management due to the complex interconnection between computing components and DM components. Since the network channels need to handle hybrid tasks, one should manage the network channel carefully to maintain data consistency and address the bandwidth competition problem. 

\textbf{Computing and memory networks (CMN):} The computing component often accesses far memory through intra-server I/O fabrics on the same mainboard. The most commonly used fabric is PCIe, which is used on both CXL memory  \cite{cxl,pond-cxl} and SSD  \cite{CXLbasedSSD,asplos19software,tmo}. 
The computing component can also access memory resources of the pass-through server through computing and memory networks, as shown in Table \ref{tab:cluster-architecture}-a. This network is responsible for far memory data transfer with remote servers. Commonly used devices include TCP/IP NIC Switch   \cite{rdmarpc-sigcomm,grpc}, RDMA NIC switch   \cite{fastswap,infiniswap,xmempod,hyfarm-iccd,buddycompression}, FPGA switch   \cite{Thymesisflow}, CXL PCIe switch   \cite{beacon-micro22,mind,memtrade}.

Thymesisflow  \cite{Thymesisflow} and Memtrade  \cite{memtrade}  are mainly based on CMN. Thymesisflow   \cite{Thymesisflow} features two switchable roles: computing role and memory-stealing role. The computing role introduces remote memory to a physical address space range; the memory-stealing role reserves a portion of the local memory on the host system, which is exposed as disaggregated memory to a neighbor host. Memtrade  \cite{memtrade} adopts a producer-and-consumer model that allows memory producers to lease both their
unallocated memory and allocated-but-idle application memory to remote consumers for a limited period of time. 

Looking ahead, Advancements in CMN pave the way for homogeneous architectures to provide high-performing all-hardware disaggregated memory solutions. For example, Gen-Z  \cite{genz} aims to provide a disaggregated I/O fabric. In addition to CXL  \cite{cxl}, OpenCAPI   \cite{opencapi} and NVLink2  \cite{nvlink} are also promising cache-coherent attachment technology for off-chip peripherals. 
%$ It designs the run-time attachment/detachment of byte-addressable disaggregated memory to a running Linux Kernel exploiting dynamically created NUMA nodes to host the remote memory.

\subsubsection{Heterogeneous CDM architecture}

In \textit{heterogeneous} CDM, DM resources are often decoupled with the computing components. It maintains higher flexibility of DM management due to the complete disaggregation of memory resources. The architecture also leaves data control and data consistency to memory pool. However, it may suffer a noticeable overhead of far memory management, especially when the far memory pool is large. Different from designs in homogeneous CDM architecture, one can adopt different NIC devices or isolated data transfer channels for distributed computing network (DCN) and memory access network (MAN).

\textbf{Distributed computing network (DCN):}  The computing units rely on DCN to handle data communication to support parallel designs (Table \ref{tab:cluster-architecture}-b). The DCN is mainly realized through a classical distributed computing network, responsible for computation distribution and task distribution   \cite{tensorflow-RDMA,grpc,PowerLyra}. 

\textbf{Memory access networks (MAN):} Similar to CMN, MAN handles data transfer with far memory devices (Table \ref{tab:cluster-architecture}-c). There have been works utilize RDMA to access remote DRAM    \cite{fastswap,infiniswap,xmempod}. Some prior works design programmable FPGA devices or smart NICs to access the remote DRAM   \cite{Thymesisflow,clio,memliner}. Recent works are more focused on utilizing PCIe-based CXL switches (CXL 2.0) to access remote CXL memory or SSDs directly   \cite{cxl, CXLbasedSSD, beacon-micro22}. Pond simulates CXL as a NUMA node without CPU cores and designs a CXL-based full-stack memory pool with an accurate memory allocation prediction model  \cite{pond-cxl}. The CXL-ANNS enables highly scalable approximate nearest neighbor search (ANNS) services by disaggregating DRAM via CXL and placing all essential datasets into its memory pool  \cite{CXL-ANNS}. Yang et al.   \cite{CXLbasedSSD} explores the design space of flash devices instead of memory devices on new CXL technologies to overcome the memory wall.

\textbf{Memory network (MN):} The disaggregated memory pool is organized by the MN, which is often based on on-chip design with high bandwidth and high memory density   \cite{stringfigure, hybrid2} (Table \ref{tab:cluster-architecture}-d). Many works design large memory nodes based on memory networks. Kim et al. propose a distributor-based network on HMC to reduce the network diameter while properly distributing the bandwidth across different routers  \cite{memorynetwork1}. Zhan et al. propose a DRAM-based unified memory network architecture on  Hybrid Memory Cubes (HMC) that provide an intelligent I/O interface to reuse the intra-memory NoC as the network switches for inter-memory communication  \cite{memorynetwork2,themachine}. Poremba et al.   \cite{memory-network} observe that placing NVM cubes in a specific order in the MN improves performance by reducing the network size/diameter up to a certain NVM to DRAM ratio.

%% file: Contents/4-System-adaption.tex
\section{System-level Adaption and Design} \label{sec-system}
On top of disaggregated memory architectures, runtime design and optimization strategies for far memory access is crucial. They provide the necessary support for high-performance far memory access and efficient memory resource scheduling.
Based on the above architecture-level design, this section summarizes the software environment of disaggregated memory, including system-level OS adaption, runtime design and task scheduling. These systems are designed to support highly flexible resource usage, low-overhead data processing, and high task/data throughput. %All these system designs provide a more flexible way for the upper-level application to run in a user-friendly way and achieve higher performance. 

\subsection{Operating System Adaption}\label{subsec-os-design}
The OS-level design leaves all the control to the OS without application-specific adaptions and thus is suitable for any general-purpose programs. One can add new control modules into the OS to manage the DM, detailed in Sec. \ref{subsec-addition-control}. One can also modify the existing modules to support the now far memory access path, detailed in Sec. \ref{subsec-modification}.

\subsubsection{Addition of control modules}\label{subsec-addition-control}
Adding new control modules into the OS can manage the disaggregated memory devices well. We categorize related methods into three groups, including the separate components control, separated power control, and separated memory control, as shown in Table \ref{tab:addition-module}.

\begin{table}[ht]\small
\centering 
\caption{Different ways of control module addition.}
\label{tab:addition-module}
\begin{tabular}{|c|c|c|c|}
% \begin{tabular}{|m{1cm}|m{4cm}|m{4cm}|m{4cm}|}
\hline
\textbf{Type} & Separated components control & Separated power control & Separated memory control \\
\hline
\textbf{Arch.} & 
\includegraphics[height=3cm]{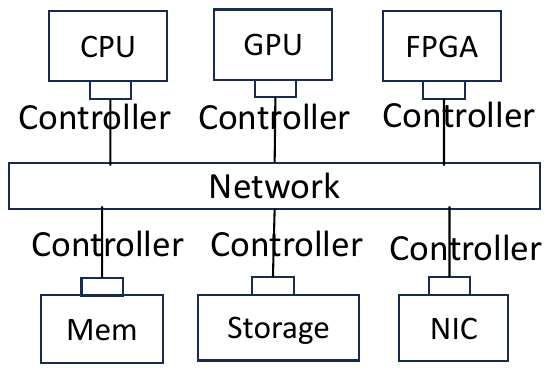} & \includegraphics[height=3.4cm]{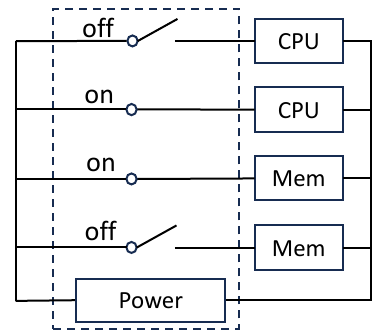} & \includegraphics[height=3.5cm]{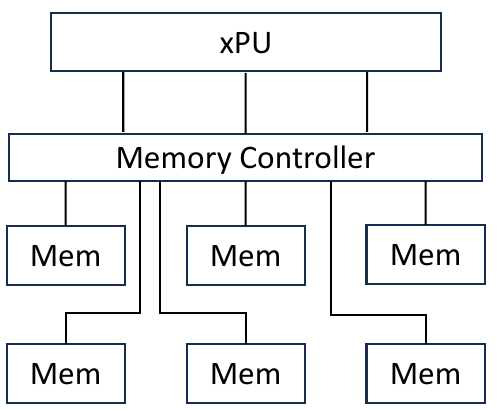} \\
\hline
\textbf{Ref.} &\makecell[l]{TCP/IP network: LegoOS \cite{legoos},\\ RDMA network: Teleport \cite{teleport-sigmod}} & Zombieland  \cite{zombieland} & \makecell[l]{  Memory chip: String Figure  \cite{stringfigure}, \\ CXL controller: BEACON  \cite{beacon-micro22}, \\ Switch controller: Mind  \cite{mind},\\ FPGA controller: ThymesisFlow \cite{Thymesisflow}, \\ I/O controller: Gen-Z \cite{genz}} \\
\hline
\end{tabular}
\end{table}

%每一个工作一句话，只讲跟主题有关的部分

\textbf{Separated components control:} 
A basic approach is to use isolated control subsystem for separated CPU, memory, network, and storage. Each device monitor can operate locally for its functionality and only communicates with other monitors for real-time resource requirements. LegoOS \cite{legoos} introduces a modular system named \textit{splitkernel} to decouple the traditional OS functionalities into loosely coupled monitors, each of which runs on and manages a hardware component. Each control module has only two global tasks: orchestrating resource allocation across components and handling component failure. 

%One can also design a framework to reduce the potential overhead of network latency into every load and store operation \cite{teleport-sigmod}. They also describe a set of specialized synchronization primitives that guarantees memory coherence of a logical process context shared across resource pools with multiple concurrent threads handling. 

\textbf{Separated power control:}  
Energy costs can be saved if the power of CPU and memory are managed separately and provisioned on demand. Zombieland system \cite{zombieland} provides an effortless way for disaggregating the CPU-memory couple at the power supply domain level. CPU and memory still share the same board, but their power supply domains are separated. It also uses a new ACPI sleep state (called zombie), which allows the suspension of a server (thus saving energy) while making its memory remotely accessible. In addition, the consumed memory node is equipped with specialized hardware controllers with separated power control  \cite{clio,mind}.

%for minimalist low-power processors to handle rarer control plane operations???

\textbf{Separated memory control:} 
Independent control modules can manage memory devices, handling data read and write on memory blocks and the data coherency of the memory pool. At the hardwre level, a classic solution is to build a sizeable on-chip memory pool organized with routers and managed by a global memory controller. For example, String Figure  \cite{stringfigure} proposes a high-throughput, elastic, and scalable memory network architecture. It generates random topology for high network throughput and near-optimal path lengths in large-scale memory networks. At the software level, one can use appropriate software to manage memory resources in both memory blades and switches. For example, BEACON  \cite{beacon-micro22} is located near the memory pool and it can process data in the memory pool and accelerate data routing in the switch. This design can leverage the high communication bandwidth provided by CXL without modifying cost-sensitive DRAM. Further, there are several works adopting hardware-software co-design ways to manage the memory pool. For example, ThymesisFlow \cite{Thymesisflow} introduces a software-defined hardware/software co-designed datapath on top of the IBM POWER9 \cite{power9}. It embraces the OpenCAPI  \cite{opencapi} cache coherent attachment technology that is available today on IBM POWER9 processors \cite{power9}. It uses a Remote Memory Management Unit (RMMU) to manage the address translation and make each memory section (viewed as far memory space) independently and "hotplugged" at runtime. Clio \cite{clio} proposes a hardware-based memory disaggregation solution that has the right amount of processing power at memory nodes. Implemented on FPGA hardware, Clio virtualizes and manages disaggregated memory at the memory node, including a new hardware-based virtual memory system, a customized network system, and a framework for computation offloading. 

%However, proposing completely new types of operating systems may bring new problems. On the one hand, they altogether abandon the original task execution process and cannot be optimized for general-purpose tasks with traditional operating systems that take advantage of continuous iterations. On the other hand, they require entirely new hardware support, making it challenging to deploy them quickly into existing software systems.

%It also employs a mix of computation and lookup tables to reduce routing overhead routing. It prepares a set of network reconfiguration mechanisms that allow both static and dynamic network expansion and reduction. 

%This enables efficient on-demand memory expansion with unmodified CXL-DIMMs and eliminates the performance bottleneck of communication. 

% With the proposed architecture design (e.g., processing in the memory pool and in-switch data routing) and memory management framework (e.g., memory allocation and address mapping), BEACON enables efficient on-demand memory expansion with unmodified CXL-DIMMs and eliminates the performance bottleneck of communication. BEACON-D is a Processing-In-DIMM accelerator and performs the computation within the enhanced CXL-DIMMs. As a comparison, BEACON-S is a Processing-In-Switch accelerator and performs the computation within the enhanced CXLSwitches. Both BEACON-D and BEACON-S maintain the non-invasive feature and practicality of the previous DIMM-based accelerators for genome analysis without making any modification to the cost-sensitive DRAM dies.

\subsubsection{Modification of existing modules}\label{subsec-modification}
Many prior works propose new technologies at the software level to fully use existing mechanisms in the OS. These designs can be quickly adapted to deploy on commercial hardware architectures while ensuring transparency to the task. We summarize prior works into three categories, including network card replacement, storage replacement, and memory replacement, as shown in Table \ref{tab:modification}. Traditionally, the swap space is a block partition on the disk, which is managed as a file. In recent years, researchers have replaced the original swap backend with higher-performance memory devices to optimize the data swap performance, such as RDMA network cards  \cite{infiniswap,xmempod,fastswap,icdcs19} and SSD \cite{xmempod,tmo,asplos19software}. A basic implementation involves adding a kernel patch and a new module into the OS, utilizing data transfer procedures with far memory access.

%The advantage is that one only needs to limit the use of local resources through the operating system adaptation, including cgroups, etc., to trigger the swap mechanism with far memory calls. 

\begin{table}[t]\small
\centering 
\caption{Differerent ways of existing module modifications. }
\label{tab:modification}
 \begin{tabular}{|c|c|c|c|}
% \begin{tabular}{|m{1cm}|m{4.5cm}|m{4.5cm}|m{4.5cm}|}
\hline
\textbf{Type} & Network card replacement & Storage replacement & Memory replacement \\
\hline
\textbf{Arch.} & 
\includegraphics[scale=0.6]{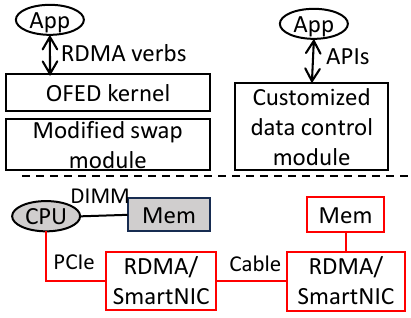} & \includegraphics[scale=0.6]{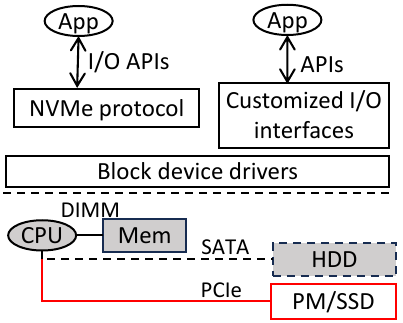} & \includegraphics[scale=0.6]{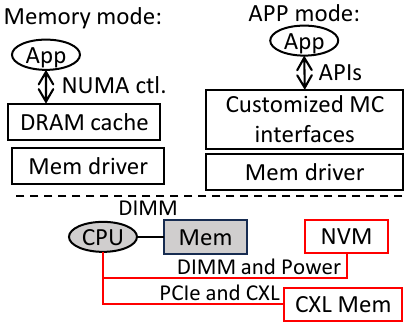} \\
\hline
\textbf{Ref.} & \makecell[l]{TCP/IP card:  \cite{grpc},  \cite{liang2005swapping}, etc. \\ RDMA:  \cite{infiniswap},  \cite{fastswap}, etc. \\ SmatNIC:  \cite{clio},  \cite{Thymesisflow}, etc. \\ Smart switch:  \cite{mind} etc. } & \makecell[l]{Disk and SSD:  \cite{asplos19software}, etc. \\Heterogeneous SSDs:  \cite{tmo}, etc.\\ SSD and RDMA:  \cite{xmempod},  \cite{hyfarm-iccd},etc.\\ CXL-based SSD:  \cite{CXLbasedSSD} } & \makecell[l]{NVM:  \cite{pDPM-persistentmemory},  \cite{dpm-shanyizhou}, etc.\\ CXL: \cite{pond-cxl},  \cite{directcxl}, etc. \\Optical:  \cite{optically-jpdc2022,opticalFiber} ,etc.} \\
\hline
\end{tabular}
\end{table}

%\textbf{Background: Modification of data swap module.} 不是tutorial!!!
%Using disaggregated memory requires the data swap between local and far memory. At the OS level, this is often handled by the \textit{page swap} mechanism. The page swap mechanism is responsible for offloading pages when memory reaches limitations and refining and loading pages as soon as required. According to the LRU strategy, the swap mechanism reclaims the memory space by eviting the pages accessed relatively infrequently, i.e., cold pages. The file-backed pages can be dropped immediately since they have backups on the storage; the anonymous pages should be swapped out instead of deleted without backups. The swapped-out page data are stored in the far memory space with a cache indexed in the local swap frontend. Once a to-be-accessed page is not in the local memory, a page fault is triggered, and the page is fetched from far memory space to the local memory.

%Typically, memory accesses are in pages (4KB), which are swapped to storage-like disks using the Linux kernel's swap mechanism. The page swap mechanism relies on a frontswap \cite{frontswap} and a swap backend \cite{zswapkernel}. The swap front stores the pages to be swapped and the swap backend is connected to a swap space that stores the offloaded data of the application. 

\textbf{Network card replacement:} Many works have been devoted to adapting page swap mechanisms in existing systems to RDMA-based far-memory environments, as shown in Table \ref{tab:modification}. For example, Liang et al.  \cite{liang2005swapping} designs a high-performance networking block device (HPBD) over InfiniBand fabric, which serves as a swap device for efficient page transfer to/from remote memory servers. Infiniswap  \cite{infiniswap} is the first far memory system based on a traditional page-swap mechanism that utilizes RDMA as swap backends. Cao et al. \cite{icdcs19,xmempod} designed the FastSwap tool to efficient page swapping via RDMA in VMs. FastSwap can leverage idle host memory and redirect the VM swapping traffic to the host-guest compressed shared memory swap area. Fastswap \cite{fastswap} explores the role of RDMA-based far memory in enhancing task throughput.  Sherman \cite{wang2022sherman} introduces a write-optimized B+ Tree index for disaggregated memory to boost the write performance of RDMA-based far memory access. 
    
\textbf{Storage replacement:} Commercial cloud-native corporations often use storage to offload data due to the high cost of RDMA deployment, as shown in the third column of Table \ref{tab:modification}.
Google in its work Zswap \cite{asplos19software} showed that software-defined far-memory systems can actively compress cold memory pages in the DRAM. Meta's data centers use TMO system  \cite{tmo} to offload task data to far memory in heterogeneous datacenter environments. It designs a Pressure Stall Information (PSI) kernel to automatically adjust how much memory to offload to heterogeneous devices according to the device’s performance characteristics and the application’s sensitivity to memory-access slowdown. Koutsovasilis et al. \cite{holistic2021ccgrid} presented a memory balancing policy that autonomously migrates memory pages across local and disaggregated memory.  Xmempod  \cite{xmempod} uses different software managers to support data swap on inter-VM DRAM-based far memory, RDMA-based far memory, and block-device-based far memory. Hyfarm  \cite{hyfarm-iccd} utilizes both SSD-based intra-server far memory and RDMA-based inter-server far memory that replaces the disk-based swap backend and thus balances the memory usage between servers well. Several works have utilized persistent memory as a remote storage space to expand memory resources further, taking advantage of its much better performance benefits and persistence over disk to store data  \cite{dpm-octopus,dpm-shanyizhou}. Clover \cite{pDPM-persistentmemory} proposes to separate persistent memory and connect them to compute nodes in a remote storage fashion, allowing all compute nodes to directly access and manage storage nodes. 
% \textbf{2). Modification of block device access module. }

%A characteristic of persistent memory NVM is the two modes. The memory mode utilizes the original DRAM as cache and no persistency, and the APP direct mode has no more local memory space with persistency. 
 
\textbf{Memory replacement:} Recent works also utilize volatile and non-volatile memory devices as far memory space, as shown in the fourth column of Table \ref{tab:modification}. Data units are often called objects with variable data sizes, which can be stored on NVM, CXL, and optical memory devices. A key idea is to design data indexing methods based on specific efficient data structures since the data offloading and fetching in the DM system is similar to traditional key-value store operations. For example, classic DM indexing works  \cite{wang2022sherman,btree-sigmod} use B+ tree to build range indexes. dLSM \cite{icde-dlsm-disaggregatedmem,vldb-lsm-disaggregated}  proposes optimized LSM-tree compared with the traditional B-tree indexing for disaggregated memory, leveraging near-data computing with RDMA-specific customizations and tuning in byte address granularity with high system performance. Skadi  \cite{skadi-huawei} and SMART \cite{smart-huawei} argue that radix tree is more suitable for DM than the B+ tree due to smaller read and write amplifications. They design high-performance locks for the leaf nodes and read-delegation and write-combining techniques to reduce redundant I/Os.

\subsection{Runtime Design and Optimization}\label{subsec-virtualized}
Far memory systems for virtual environments are designed for better runtime management. 
Different from designs at the OS level, virtualization-level design is software-defined and non-transparent for applications. They often maintain more flexible memory management strategies, including memory reclamation, memory allocation, and memory calls.

Virtualized environments are adapting to more far-memory devices, which will be one of the main directions for research on disaggregated memory systems in the following decades. The virtualization method includes bare-hardware virtualization, virtual machines with inner OSs,  container-based virtualization, and Java virtual machines to provide isolated resources.

%The runtime optimization of the virtual layer provides an object-based management abstraction.

\subsubsection{Memory and network virtualization}
%在虚拟化环境中，人们可以通过内存地址映射和网络通道隔离来虚拟化内存和网络资源。对于内存虚拟化，系统建立虚拟地址到物理地址的复杂映射来支持VM内部到Host的灵活内存管理。通常情况下，内存以多个单独的内存条存在，根据它们和CPU socket的距离可以划分不同的内存区域，即NUMA架构。同一个socket中的内存访问性能相同，但跨socket的内存访问会有更低的带宽和更长的延时。

% \begin{figure*}[t]
%     \centering
%     \includegraphics[width=0.9\linewidth]{figure/numa-and-network.pdf}
%     \caption{The virtualization environment considers NUMA architecture and network channel features to better isolate resources. }
%     \label{fig:numa-and-network}
% \end{figure*}

In a virtualization environment, memory and network resources can be abstracted through memory address mapping, memory region isolation, and network channel allocation.
\begin{itemize}
    \item \textit{Memory address mapping}: For memory virtualization, the system establishes a complex mapping of virtual addresses to physical addresses to support flexible memory management from inside the VM to the Host. Works \cite{mind} can build a global virtual address space shared by all processes and range partitioned across memory blades to minimize the number of address translation entries.
    \item \textit{Memory region isolation}: Typically, memory exists as multiple individual memory sticks that can be divided into different memory regions based on their distance from the CPU socket, i.e., NUMA architecture \cite{cpu-less-numa}. Memory accesses in the same socket have the same performance, but memory accesses across sockets will have lower bandwidth and longer latency.%, as shown in Figure \ref{fig:numa-and-network}-(1).
    \item \textit{Lightweight network channel allocation:} On general TCP/IP and RDMA network cards, the network channels are often controlled by the CPU cores since they are bonded with CPU threads.
    %, as shown in Figure \ref{fig:numa-and-network}-(b). 
    On RDMA, the Single-Root I/O Virtualization (SR-IOV)  \cite{rdma-vldb} is a standard mechanism of PCIe device virtualization technology. Each virtual PCIe device Z(such as RDMA) has its own PCIe configuration space and provides services for upper-layer software just like physical PCIe devices. 
    \item \textit{Easy-to-use interfaces abstraction}: people have designed diverse far memory access interfaces to easy use RDMA-based transfer\cite{farm-nsdi14,lite,remote-regions,freeflow}. A proper implicit data structure is used behind the data transfer on memory spaces across network.  
\end{itemize}
 
% \textbf{Implementation}: Existing works often adopt OS-supported tools to control the CPU core, memory usage, network channel, swap space, and storage partitions for each process. Generally, existing works often utilize \textit{Cgroup} \cite{cgroup} and \textit{namespace} \cite{namespace} to limit the CPU core usage and memory usage bound. 
% One can use QEMU \cite{qemu} in KVM \cite{KVM} to manage the created virtual far memory paths in virtual machines.

%We then summarize the user-defined resource isolation into three types. The first is local memory isolation. Existing works \cite{fastswap,fargraph,tmo} often adopt OS-supported tools to control the CPU core, memory usage, network channel, swap space, and storage partitions for each process. Generally, they often utilize \textit{Cgroup} \cite{cgroup} and \textit{namespace} \cite{namespace} to limit the CPU core usage and memory usage bound.  The second is data access channel isolation. Some works \cite{infiniswap,canvas-nsdi23} build isolated data swap channels on RDMA by adding an adaptive swap allocator, which allows each application to possess its dedicated swap partition, swap cache, prefetcher, and RDMA bandwidth. The last is far memory isolation on the memory node. Related works  \cite{mind,aifm} feature domain-based memory protection with fine-grained and flexible protection. 

\subsubsection{Design in virtual machine (VM)}
%对于VM的远内存访问主要由VM申请，Host分配。Host可以分本机回收的内存，在新的基于网络的分离式环境中也可以分配位于其他机器的远内存。1）swap隔离 2）内存回收和分配。

In VMs with host OS and without host OS, there are significant differences on far memory allocation ways. Each VM will be assigned CPU and memory in advance, while a hypervisor or monitor mainly manages storage I/O and network management. In scenarios with host OS, such as VMware \cite{vmware}, KVM \cite{KVM}, etc., the local memory can be allocated to each VM by the host, and the VMs can access different remote memories, both from other VMs and by exchanging data with the host memory. In scenarios without host OS, e.g., Xen, etc., the storage I/O, and network of each VM are managed by a particular VM (also called VM 0). In these scenarios, QEMU \cite{qemu} is the core component for I/O and network management. 

The disaggregated memory devices described in this paper include memory-like \cite{cxl,pond-cxl}, storage-like  \cite{tmo,persistentmemory,CXLbasedSSD} and network-involved \cite{infiniswap,fastswap,mind,canvas-nsdi23} memory nodes. Thus, the design of a VM-level far memory system will consider the memory interactions between VMs, memory reclamation for each VM, and I/O and storage management of each VM.  The host can allocate memory already reclaimed from VMs as well as additional memory that has not yet been allocated. In a network-based detached environment, the host can also allocate far memory for VMs on other machines.

%VM far-memory accesses are mainly requested by VMs and allocated by the host or hypervisor.
% They often use QEMU \cite{qemu} in KVM \cite{KVM} to manage the virtual far memory paths in virtual machines. 

Optimizing the memory access hierarchy in a virtual machine environment can improve the efficiency of memory resource allocation. Existing works concentrate on host swapping and ballooning for memory consolidation and over-commitment. VSwapper \cite{vswapper} addresses the challenge of poor performance caused by various types of superfluous swap operations, ruined swap files sequentially, and uninformed prefetch decisions upon page faults. It provides a Swap Mapper to monitor the disk I/O mapping with memory and a false read preventer to eliminate false reads. HybridSwap \cite{hybridswap} addresses the problem that guest OS cannot utilize free pages in the host directly. They propose a distributed scalable framework to organize surplus memory in all hosts into virtual pools for swapping. XmemPod \cite{xmempod}  can dynamically expand the memory capacity hierarchically memory by expanding its memory demand over virtualized host memory first and remote memory next before resorting to an external disk.  Pond \cite{pond-cxl} can create machine learning models that can accurately predict local and pool memory allocation rates on VMs with same-NUMA-node memory performance.

\subsubsection{Design in containers} In the cloud, far memory management in containers is attracting more attention, such as the main services Microservices and Serverless functions.
There are optimization methods designed for containerized clouds.
Some works optimize RDMA-0based far memory access in container environments.
Freeflow \cite{kim2019freeflow} virtualizes RDMA environment for containerized clouds. It supports multi-tenancy isolation, container portability, and control and data plane controllability. Mitosis \cite{mitosis-serverless} proposes a fast RDMA remote fork that can quickly launch containers on multiple remote machines. 
% \textbf{Serverless function optimizations on far memory: }
% % Serverless architectures allow users to build and run services without managing the underlying infrastructure by offering on-demand computing and storage resources.
Some works study the disadvantages and overheads when deploying \textit{serverless} functions. Medes  \cite{saxena2022memory} optimizes the system performance by identifying memory redundancy and designing a simple sandbox management policy that supports smooth navigation in the trade-off space for operators. Unlike prior systems that allocate memory at job granularity, Jiffy  \cite{jiffy-eurosys22} can achieve task-level resource scheduling. Jiffy efficiently multiplexes memory capacity across concurrently running jobs, reducing the overheads of reads and writes to slower persistent storage.  Skadi  \cite{skadi-huawei} is a distributed runtime to mitigate the failure of deploying resource disaggregation in a harmony system. Beldi  \cite{zhang2020fault} proposes a stateful serverless runtime system for writing and composing fault-tolerant and transactional stateful serverless functions to guarantee exactly once semantics. 
%Beldi extends Olive, a recent frame-work that exposes an elegant abstraction to clients of cloud storage systems, and adapts its mechanisims to the stateful serverless functions in order to guarantee exactly-once semantics to workflows of SSFs. Beldi also adapts existing concurrency control and distributed commit protocols to support transactions over SSF workflows.

\subsubsection{Design in Java virtual machine (JVM)}
%在java环境提供了更灵活的内存卸载和读取策略，一些工作关注JVM系统中的统一的内存管理：

Some work focuses on invocating far-memory resources in Java virtualized environments, designing strategies for their synergy with Java garbage collection (GC) mechanisms. For example, MemLiner \cite{memliner} proposes a runtime technique that “lines up” memory accesses from the application and the GC, thereby reducing the local-memory working set and improving remote-memory prefetching.  Semeru \cite{semeru} designs a distributed JVM that provides a unified abstraction of virtual memory across CPU and memory servers and a distributed GC that offloads object tracing to memory servers. 
Some work has been devoted to building application non-transparent far-memory frameworks based on custom APIs to support far-memory calls in virtualized environments.  AIFM  \cite{aifm} proposes application-integrated far memory based on remoteable and hybrid near/far memory data structures, which makes far memory available to applications through a simple API and without read-and-write amplification.

\subsection{Task Scheduling Design} 
The above runtime systems can support various tasks in the virtualized environment, where tasks share the computing, memory, network, and storage resources of servers in the datacenter. Many works design and optimize task scheduling to better utilize disaggregated resources with high flexibility and high quality of service (QoS). 

\subsubsection{DM scheduling architecture}
\begin{figure}[t]
    \centering
    \includegraphics[width=\linewidth]{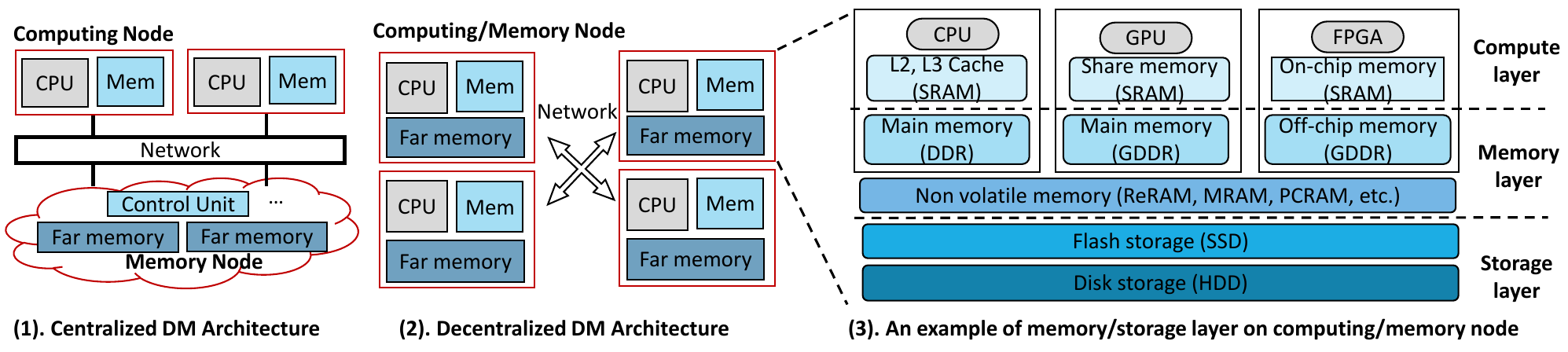}
    \caption{The centralized and decentralized DM  management method.}
    \label{fig:centralized}
    \vspace{-0.6cm}
\end{figure}

There are two types of management methods on disaggregated memory in the clusters, which we termed centralized management architecture and decentralized management architecture, as shown in Figure \ref{fig:centralized}. The centralized DM management is easy for resource management and low-cost memory updating, while the decentralized DM management performs better for resource sharing and cost efficiency.

\textbf{Centralized DM scheduling architecture: }The centralized DM scheduling   \cite{stringfigure,optically-jpdc2022,optoelectronic-memory} maintains a central management entity, such as a large memory pool with multiple memory nodes, to provide far memory resources to each computing node.
Actually, this concept was not proposed first; earlier in 1989, Kai Li et al.   \cite{1989memorycoherence} stated that, in the shared memory system, one can use a shared virtual memory as the memory management monitor. The monitor consists of a data structure  \cite{FarDataStructure} and some procedures that provide mutually exclusive access to the data structure. 
Centralized DM has the advantage of data coherence management, which the memory pool monitor can strongly protect. However, it faces the problem of scaling up and out. One can find it hard to build a larger memory pool of more than 1000 memory nodes  \cite{stringfigure}.

\textbf{Decentralized DM scheduling architecture: }
In decentralized DM scheduling   \cite{xmempod,Thymesisflow, mind}, the memory resource is distributed physically and shared virtually. Each server node is equipped with far memory resources and can provide additional memory resources for its own computing units and other computing nodes. As shown in Figure \ref{fig:centralized}-(3), each server can have SRAM-based caches, DDR DRAMs, RAM-based non-volatile memories, flash-based SSDs, and disk to hierarchically store the data for heterogeneous computing units.  This pattern can not only solve the scalability problem of centralized DM but also alleviate the performance bottleneck. Building a decentralized data communication scheme can decouple memory control and reduce data synchronization traffic. For example, COARSE  \cite{coarse-hpca22}  is a disaggregated memory extension that improves communication performance of GPUs. It builds on modern cache-coherent interconnect (CCI) protocols and MPI-like collective communication for synchronization, allowing low-latency and parallel access to training data and model parameters shared among worker GPUs. Memtrade  \cite{memtrade} employs a central coordinator that manages the disaggregated memory market and matches memory producers and memory consumers based on their supply and demand.

%Decentralized DM Architecture is to "distribute" the memory pool resource and manage the memory resource in a virtualized way. 
%

% \textbf{Memory layers on each node:} As shown in Figure \ref{fig:centralized}-(3), we further give the introduction of memory layers on each memory node. Each computing unit has its on-chip SRAMs as caches for fast calculation. The memory layer also maintains DDR-based DRAMs and RAM-based non-volatile memory devices, which are always managed to serve for fast computation. The storage layer includes flash-based storage like SSDs and disk-based storage, often acting as a large space to store data or serving as data backups. 
%SOTA shared-state schedulers\cite{omega,apollo}
In cluster-level DM architectures, homogeneous and heterogeneous architectures have their advantages and disadvantages. Both can be used for resource scheduling and management through centralized and decentralized methods. Just like the shared-stade scheduling architecture  \cite{omega,rminer-socc,apollo}, homogeneous and heterogeneous architectures may reasonably coexist and work in complementary ways in future disaggregated clusters. 

\subsubsection{ Resource management strategies} Many works optimize the system from the view of hardware resources.
We list several commonly used optimization methods for efficient resource management.

\textbf{Adaptive data distribution:}
Optimizing resource allocation and access patterns in disaggregated memory systems requires an awareness of applications' or data types' latency sensitivity. Identifying various sensitivities helps the system optimize memory access and bandwidth allocation.
HyFarM \cite{hyfarm-iccd}, CFM \cite{fastswap}, and TMO \cite{tmo} both focus on profiling and measuring the impact of far memory on application performance. CFM \cite{fastswap} develops a degradation profile for each application by measuring runtime across different local memory ratios. TMO \cite{tmo}, on the other hand, leverages the Pressure Stall Information (PSI) mechanism to dynamically adjust offloading decisions. HyFarM \cite{hyfarm-iccd} senses sensitivity to collocate FM-sensitive and FM-tolerant tasks. Comparatively, while fastswap \cite{fastswap} and TMO \cite{tmo} dive into performance degradation of individual tasks, HyFarM \cite{hyfarm-iccd} applies the sensitivity to multi-task orchestration in a hybrid FM environment.

% fastswap \cite{fastswap} profiles the performance degradation applications experience when they trade local memory for far memory. For each application, fastswap creates a degradation profile that estimates the runtime at different local memory ratios. To create a degradation profile, fastswap measures the application’s runtime at several discrete local memory ratios using Fastswap, and then uses polynomial fitting to create a continuous function that the memory policy uses 

% TMO \cite{tmo} introduces a new kernel mechanism called Pressure Stall Information (PSI), which directly measures in real time the lost work due to resource shortage across CPU, memory, and I/O. PSI is reported on a per-process and per-container basis. Unlike g-swap’s promotion-rate metric, PSI accounts for both the performance characteristics of the slow memory tier and the application’s sensitivity to memory-access slowdown. A userspace agent called Senpai uses the PSI metrics to dynamically decide how much memory to offload without prior application knowledge while taking into account hardware heterogeneity in data centers.

% HyFarM \cite{hyfarm} is a high-performance task orchestration scheme for hybrid FM. It prudently collocates FM-sensitive and FMtolerant tasks together for better efficiency. Importantly, it develops a performance optimization approach that combines both memory expansion and extension, which greatly improves memory utilization and performance per bit.

% Direct memory/storege access can skip I/O access overheads as well as reduce the occupation of local memory space.
\textbf{I/O bandwidth expansion:} 
   Existing works try to add new direct memory/storage access paths to improve the memory efficiency and I/O bandwidth utilization. GPU direct storage (GDS) supports direct storage that skips copies in CPU memory. However, the parallelization of GDS's I/O requests is limited by the CPU. BAM framework \cite{bam} adds NVMe-based I/O read and write logic to the cuda core, enabling the GPU to directly initiate more threads. The work GIDS  \cite{GIDS} further adds asynchronous I/O processing modes based on BAM, thus further reducing I/O overheads. Direct memory access is also a solution to reduce data copy and duplication. For example, Zero-Infinity  \cite{Zero-infinity} can offload all of the partitioned model states to CPU or NVMe memory or keep them on the GPU based on memory requirements, thus handling huge parameters for training on current GPU clusters. 

% which is able to initiate I/O requests directly through the GPU, using only the CPU's I/O request logic without reading or writing data through the CPU
%In addition, reducing the occupation of memory space is a crucial objective of direct memory access. 
% \subsubsection{?High bandwidth for data transfer} 
% High bandwidth is crucial for supporting fast data transfer between compute nodes and memory pools.

% Canvas \cite{canvas} isolates swap cache, swap partition, and RDMA bandwidth to prevent applications from invading each other’s resources. Now that resource accounting is done separately for applications, Canvas offers three optimizations that adapt kernel operations such as swap-entry allocation, prefetching, and RDMA scheduling to each application’s resource usage, providing additional performance boosts.

\textbf{Data throughput improvement:}
The nature of disaggregated memory systems supports parallel data transfer and parallel execution of applications by allowing concurrent access to a shared memory pool. 
Fargraph+ \cite{fargraphplus} proposes a parallelism control strategy on graph processing on disaggregated memory architecture. It controls the multi-threading design of graph applications on the basis of saving RDMA-related resources (CPU core, memory space, RDMA queues, etc.
Canvas \cite{canvas-nsdi23} redesigns the swap system with fully isolated swap paths for remote-memory applications, thus improving the swap parallelism of data swap paths.

\textbf{Resource preemption reduction:}
To reduce resource preemption in disaggregated memory systems, effective resource isolation among different tenants or applications is essential.
 Canvas \cite{canvas-nsdi23} isolates swap paths for applications, each processing its own swap partition, cache, and RDMA bandwidth. Remote Region \cite{remote-regions} supports exporting memory as files, supplementing RDMA functions lacking, such as namespace and access control. 
To balance the diverse needs of applications, ensuring high-priority tasks with more local memory resources is essential. HyFarM \cite{hyfarm-iccd} works hard on balancing the resource pressure between servers by analyzing the far memory sensitiveness of each application and assigning more local memory to those far-memory-sensitive tasks.

% Supporting a huge number of concurrent clients is critical for far memory environment. To ensure tight and efficient resource packing, we should allow many (e.g., thousands of) client processes running on tens of computing nodes to access and share a memory node. This scenario is especially important for new data-center trends like serverless computing and microservices \cite{cilo}. 

% The Clio \cite{clio} includes a virtual far memory system, a customized network system, and a framework for computation offloading. Cilo co-designs software with hardware, CNs with MNs, and network stack with virtual memory system to handle the entire data path of the memory nodes in hardware with high throughput, low latency, and minimal hardware resources.

% \subsubsection{OS-level Optimizations} Some works use OS-level optimizations in the OS level aiming at specific design goals.

 \textbf{Reliability guarantee:}
Existing work trends to improve the system's reliability include failure recovery, fault tolerance, resource isolation, etc.
1). Some work has been devoted to failure recovery designs to improve the reliability of far memory systems. 
Hydra \cite{Hydra} presents a failure resilience mechanism for remote memory, adopting an in-memory erasure coding scheme that achieves single-digit µs tail memory access latency by analyzing load balancing and availability trade-offs for distributed erasure codes. 
Carbink \cite{Carbink} proposes a far memory system that uses erasure-coding, remote memory compaction, one-sided RMAs, and offloadable parity calculations to achieve fast, storage-efficient fault tolerance. %Compared to Hydra, Carbink has lower tail latency, higher application performance, and higher memory usage. 
2). Some works adopt coherence protocols for better fault tolerance.
Āpta  \cite{apta-fault-tolerant-cxl} designs a fault-tolerant object-granular CXL disaggregated memory for accelerating Function-as-a-Service (FaaS). %CXL lacks the requisite level of fault tolerance necessary to operate inter-server far memory access with inefficient data store accesses. 
Apta introduces a novel fault-tolerant coherence protocol for keeping the cached objects consistent without compromising availability when facing computing node failures.
3). Plenty of works implement isolated far memory access to avoid failure and interference. For example, the vVEEs \cite{trusted-apsys23} construct trusted and virtual Trusted Execution Environments (TEEs) by adding unified interfaces and standard security primitives that can run on heterogeneous hardware components. It also builds secure domain isolation across disaggregated components at the user level without losing flexibility and elasticity. Xmempod \cite{xmempod}, Memory trades  \cite{memtrade}, and Remote reagions \cite{remote-regions} use VMs and RDMA memory regions to isolate data of each program to avoid performance interference.

\subsubsection{Far-memory-aware scheduling}
There are some works designing far memory aware scheduling methods for higher memory utilization and efficiency. 

%Ttraditional memory scheduling is an important basic for far memory resource management. Meanwhile, there are 

\textbf{Memory resource scheduling:}
The traditional scheduler needs to ensure strict Quality of Service (QoS) of tasks.  Some tasks use elastic scaling to flexibly schedule resources \cite{roy2011efficient}, improve resource management efficiency through task understanding and resource occupancy prediction \cite{cortez2017resource}.
Some works try to dynamically capture accurate resource requirements from execution runtime for fine-grained scheduling \cite{jin2020improving}, and optimizing jobs with complex dependency structures and heterogeneous resource requirements \cite{grandl2016graphene}. Others are geared towards goals such as scheduling fairness, energy efficiency, and cost control. 

%Cluster resource scheduling tasks will be roughly divided into two kinds, one delay-sensitive (latency critical, LC) task, and one batch (Best Effort Batch, BE) tasks, the scheduler usually use a mixed deployment of LC tasks and BE tasks to improve the overall resource utilization  \cite { rzadca2020autopilot,suresh2021servermore}.

\textbf{Dynamic data placement:} 
Most of the recent scheduling approaches focusing on dynamic data placement. Memtrade \cite{memtrade} harvests producer memory using an application-aware control loop to form a distributed transient remote memory pool with minimal performance impact. For example, CFM  \cite{fastswap} computes the minimum amount of data that can be retained in local memory for each task's SLO requirements, thus freeing up more memory space than is available to accommodate more tasks. Software-defined remote memory  \cite{asplos19software} compresses the swapped-out pages by the zswap kernel \cite{zswapkernel} and selectively evict low-priority jobs by killing them and rescheduling them on other machines when memory is reaching its limit. xMemPod  \cite{xmempod} employs a hierarchical remote memory system that extends the VM's memory to the VM's mainframe memory, to the remote memory, and finally to the hard disk. Allot \cite{DHMP-hust} proposes memory resource abstraction and data placement policies for distributed hybrid memory pools (DHMP) with RDMA capabilities to efficiently manage hybrid NVM and DRAM memory.

%uses simulation for cluster-level scheduling emulation???

% Fastswap \cite{fastswap} proposes a far memory-aware cluster scheduler which can place jobs on servers with insufficient local memory, ensuring that all jobs have access to sufficient total memory. This work simulates and evaluates tasks throughput improvement using far memory at rack scale.
\textbf{Netural-network-based scheduling: }Some works design hybrid memory schedulers for natural networks. 
Sentinel \cite{ren2021sentinel} states that HM imposes challenges on tensor migration and allocation for high-performance DNN training. Sentinel uses dynamic profiling and coordinates OS and runtime-level profiling to bridge the semantic gap between OS and applications, which enables tensor-level profiling and co-allocating tensors with similar lifetime and memory access frequency into the same pages. 
Kleio \cite{doudali2019kleio} proposes a page scheduler with machine intelligence for applications that execute across hybrid memory components. The scheduler combines existing, lightweight, history-based data tiering methods for hybrid memory, with novel intelligent placement decisions based on deep neural networks.

\textbf{Application-aware scheduling:}
Some works take the application sensitivity of far memory usage into account. By directly measuring an application’s sensitivity to memory access slowdown, TMO  \cite{tmo} can effectively offload memory across diverse workloads with minimal impact on application performance. CFM  \cite{fastswap} designs an application-aware cluster scheduler that leverages far memory to improve job throughput. HyFarM \cite{hyfarm-iccd} proposes a task management strategy for hybrid far memory clusters that cooperatively co-locate tasks according to the far memory sensitivity analysis. It features an intra- /inter-node joint memory adaptation approach for balancing the memory usage at both the task and server levels.

%These works use Fist-in-first-cap or last-in-first-cap strategies to allocate far memory resources, which is too simple to handle all types of tasks. 

\textbf{NUMA-aware scheduling:}
Plenty of works concentrate on NUMA architecture to optimize application performance. Pond \cite{pond-cxl} relies on the CXL interconnect standard and takes the CXL memory as a NUMA node. It exposes the pool memory to a VM’s guest OS as a zero-core virtual NUMA (zNUMA) node, i.e., a node with memory but no cores, like Linux’s CPU-less NUMA \cite{cpu-less-numa}. With the combination of hardware and systems techniques, it can achieve both same-NUMA-node memory performance and competitive cost for public cloud platforms.
ThymesisFlow \cite{Thymesisflow} designs the run-time attachment/detachment of byte-addressable disaggregated memory to a running Linux Kernel exploiting dynamically created NUMA nodes to host the remote memory.
This work  \cite{holistic2021ccgrid} utilizes Linux's NUMA memory page migration policy that automatically promotes and moves hot/cold memory pages between local and far memory.  

%The advantage of the design change the coverage of the anchor entry dynamically, reflecting the current allocation contiguity status.

% \section{ Classification of Related Works}

%  \input{Contents/table.tex}

%% file: Contents/5-Application-level.tex
\section{Application-level Optimization Methods} \label{sec-runtime}
 Faced with different applications and different far memory access patterns, we summarize general-purpose execution optimization and domain-specific system design. 

%For non-transparent DM runtime, applications realize far memory access operations by calling these interfaces directly. 

%On the one hand, some works have considered the programming difficulty of the underlying RDMA operations and designed RDMA-based far memory programming models that encapsulate the underlying RDMA operations and design relatively efficient and easy-to-use programming models  \cite{farm,lite,remoteregion,freeflow}. On the other hand, people consider the computational access characteristics of memory-hungry applications, design dedicated far memory access channels for applications, and selectively offload application data to improve overall performance. Some works provide user-defined frameworks to obtain high performance  \cite{freeflow}.

%Disaggregated memory software systems rely on multiple levels of runtime design and optimization strategies.  

\begin{table}[ht]
\centering 
\caption{Data management policy: inclusive and exclusive.}
\label{tab:datarelationship}
 \begin{tabular}{|c|c|c|}
% \begin{tabular}{|m{1cm}|m{6cm}|m{6cm}|}
\hline
\textbf{Type} & Inclusive  & Exclusive \\
\hline
\textbf{Arch.} & 
\includegraphics[height=3.5cm]{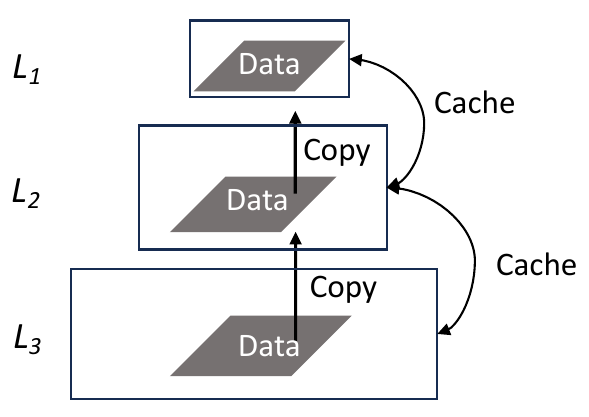} & \includegraphics[height=3.5cm]{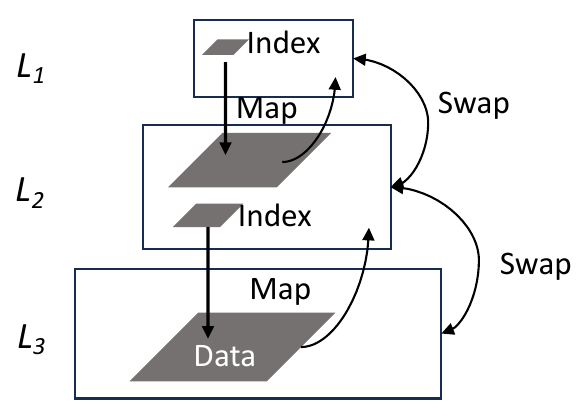}\\
\hline
\textbf{Ref.} & \makecell{ Prefetch: Leap \cite{leap},  Canvas \cite{canvas-nsdi23} etc. \\ GPU UVM: Grus \cite{wang-grus} } & \makecell{ Data map: Hybrid$^2$ \cite{hybrid2}, Sherman \cite{wang2022sherman}
\\ Data offloading: Mind \cite{mind}, Fargraph \cite{fargraph}  etc. }  \\
\hline
\end{tabular}
\end{table}

\subsection{General-Purpose Execution Optimization}
%在做任何优化设计的时候都要考虑的内容
%通常来讲，应用对于远内存的访问优化主要发生在三个流程中，1）计算节点中程序对于本地内存的访问，2）运行在计算节点的程序对本地内存和远内存的任务调度和数据卸载，即计算节点和内存节点的交互，3）内存节点上数据通过多级内存设备的缓存和存储。

% Data management for computing
% Data management for transferring
% Data management for storing

The first consideration for application performance improvement in far memory environment is the data management policy during execution. 
Particularly, the optimization strategies of application access to far memory mainly occur in three processes: 1) far-memory job distribution and local-memory data caching in compute nodes, 2) data transferring between local and far memory, i.e. data communication between compute nodes and memory nodes, and 3) data caching and storing procedure through hierarchical memory devices on memory nodes.
To simplify this, we summarize that  \textbf{FMA-based data processing = data distribution + data communication + data preservation}.

\subsubsection{Data distribution}

Far-memory-involved computation is a new type of computing pattern that is complementary to the existing computation mode. We discuss how far memory acts in program running procedures and the relationship between far memory and existing computing patterns. 

% \textbf{Computing and Memory access:}
% %任务计算访存流程，有的计算密集，有的访存密集，有的io密集
% Researchers mainly use three functions provided by virtual memory subsystem \cite{rethinking-asplos21}. 1) Using page fault to detect remote memory access and cache remote pages into local DRAM. 2) Setting the protect bit of cached remote pages to track dirty data and causing a write page fault on the first write to each cached remote page to write back dirty data. 3). Setting the present bit as not present to evict cached pages to far memory and then flushing the translation lookaside buffers to avoid accessing not-present pages.

\textbf{In-memory data offloading:} To save local memory space, large-scale applications, such as graph processing  \cite{ligra,graphit}, video processing \cite{ffmpeg}, AI model training \cite{shao-icpe, wang2018superneurons} applications, often load many tax-like few-used data into local memory. Based on fine-grained program analysis, far memory can help to offload the in-memory data into far memory, including offloading cold and read-only data \cite{fargraph,aifm}, evicting one-off data \cite{fastswap,tmo}, compressing cold data \cite{buddycompression,zswapkernel}, et al. 

\textbf{
out-of-core data distribution:} Existing works often \cite{gridgraph,cagra,zhang2018wonderland,chaos,mosaic} use classic storages (i.e. HDD) to store large dataset. Far memory systems can accelerate the I/O performance by replacing the HDD-based access paths and programming interfaces with RDMA-based\cite{fargraph,fastswap} or SSD-based\cite{tmo,icdcs19} far memory access paths, leveraging much larger data bandwidth and shorter access latency.

% In high-performance computing clusters, works study data offloading strategies to process in-storage data in batches. Far memory can accelerate this computing mode by replacing the slow I/O operations with higher-bandwidth far memory access. 
% Some works  \cite{shao-icpe, wang2018superneurons,zhang2018wonderland, gridgraph} analyze performance implications of classic memory-consuming workloads such as graph processing and AI model training, etc.

% \subsubsection{Memory oversubscribing}: There is plenty of work on running large-scale applications on memory-constrained systems. Some works design memory-saving techniques and data offloading strategies on limited local memory and far memory space \cite{fargraph,wang-grus,wang-skywalker}.

% \textbf{Single-node and distributed computation:} As a complement to the single-node and distributed computing models, far memory provides a new option for scaling out memory with high performance. With all data loaded in local memory, the traditional single-node system often shows the best performance. 
% 1) Flexible memory extension for single nodes: The key advantage of DM compared with single-node computation is to provide a much larger memory capacity for flexible memory extension, no matter on a monolithic server node\cite{fastswap} or a virtualized machine node\cite{xmempod}, alternative to disk-based memory extension. More importantly, far memory access shows much higher performance compared with disk I/O. 

\textbf{Data partition in distributed computation:} At the cluster level, distributed refers to dividing the task process and task data into separate servers and running them in parallel\cite{PowerLyra}. With far memory involved, one can only partition task data with few job distributions, which is more friend to distributed computing\cite{aifm,remote-regions,lite,grpc}. One can also provide flexible and additional memory spaces for each sub-task to enhance overall efficiency\cite{fargraphplus}. 

\textbf{Task working set partition:}
In memory processing applications load all the working set  \cite{workingset-wiki, workingset} into local main memory and then execute them in memory and caches. The working set mainly includes two parts: the required software environment and source data. In the cloud, most of the serverless functions and micro-service applications use this model. For latency-critical applications, the in-memory processing model can have high performance but occupy much more memory space, always several times the least local memory size since there is part of the memory tax  \cite{tmo}.
Fastswap \cite{fastswap} employs Linux control groups (cgroups) \cite{cgroup} for working set partition. It sets limitations on physical memory allocation to a process to control its local memory ratio. FAM-Graph \cite{fam-graph} proposes a partition scheme that separates vertex data and retrieves edge data chunks only when needed.
Another method is to compress cold data to save more time and capacity \cite{zswapkernel}. Google's software-defined far memory \cite{asplos19software} leverages zswap mechanism to compress cold pages proactively, then move them to slower storage.

% In addition, far memory offers an attractive alternative to disk-based memory extension, especially in the cloud environment \cite{asplos19software,tmo}. Far memory access shows lower performance degradation compared with disk I/O. On the other hand, faced with fluctuating workload demand, far memory represents a more convenient way to oversubscribe memory on demand.  

\subsubsection{Data communication}
The key operations of data communication between local memory (computing nodes) and far memory (memory nodes) are data offloading (to write and store data on the far memory) and data fetching (to read data from far memory and get them into local memory for computing). 
Some works design specific memory access approaches based on read and write behavior and utilize data transfer channels to improve memory bandwidth utilization \cite{cao-drdram, wang-grus}. \textit{(1). Data/Memory offloading} is a common term used to indicate that part of the data in a program can be offloaded to far memory and will be retrieved from the same memory address. Due to the limited local memory, the unused data in local memory needs to be reclaimed to free up memory space (i.e., memory reclamation or garbage collection). One can swap data from the far memory into the local memory on demand, known as memory swapping. 
%The operating system usually handles this in a fixed granularity block of data (i.e., page, usually 4 kilobytes), so it is also often referred to as page swapping. In local memory, a page fault occurs when a local program accesses data that has been offloaded to the far memory. Fault handling requires us to find this data on the far memory by a specific index and return it to the source program.
\textit{(2). Data fetching} describes retrieving data from a non-local memory. Designing an efficient fetching mechanism is necessary due to the limited memory or I/O bandwidth and high demand for access speed. One can leverage a memory management method along with a network communication protocol to fetch the remote data. 

\textbf{User-friendly program interfaces.} 
Program interfaces significantly simplify access to disaggregated memory systems, utilizing RDMA and other technologies. 
LITE \cite{lite} and FaRM \cite{farm-nsdi14} both abstract RDMA to facilitate application-level interactions. LITE \cite{lite} virtualizes native RDMA into a local indirection tier to achieve both safety and performance. FaRM \cite{farm-nsdi14} uses RDMA to build a memory-distributed platform in which address space is shared. Remote Region \cite{remote-regions} and AIFM \cite{aifm} both offer concise APIs for memory manipulation, making it easier to manage remote and local data. 
Besides, FreeFlow \cite{freeflow}, AR-gRPC \cite{tensorflow-RDMA}, and Mitosis \cite{mitosis-serverless} support high-performance applications by providing far-memory access frameworks for specific scenarios. 

\textbf{Adaptive data granularity:}
Data granularity refers to the level or size of data division. The way data is stored and processed across various hardware and software platforms significantly influences efficiency. Some works try to design smaller or more flexible data access granularity for applications to improve application performance. 
Fargraph \cite{fargraph} transparently segment data into data chunks in RDMA transfer to facilitate RDMA read/write.
Increasing data granularity to enable accurate operations is also a trend. AIFM \cite{aifm} utilizes far memory at object granularity and manages local offloadable memory at log granularity.
Kona \cite{rethinking-asplos21}  uses cache coherence instead of virtual memory for tracking applications' memory access behavior transparently at cache-line granularity. 
 Some work addresses the problem of high dirty data amplification that comes from using page granularity for tracking changes to the cached data (4KB or higher). Remote regions \cite{remote-regions} consider the trade-off between granularity and flexibility in its region map, which tracks the location of regions.

\textbf{Parallel data swap:}
The coexistence of multiple applications can lead to significant performance degradation due to interference among swap operations. This challenge underscores the importance of developing robust parallel data swap mechanisms that can accommodate the simultaneous demands of diverse applications while minimizing the adverse impacts of such interference. 
Recent works aim to optimize swap efficiency and minimize slowdowns due to cross-application interactions. Canvas \cite{canvas-nsdi23} introduces a fully isolated swap environment for each application, encompassing resources such as swap partitions, caches, prefetchers, and RDMA bandwidth. They design dynamic swap entry allocation, semantics-aware prefetching, and strategic RDMA scheduling, collectively enhancing swap efficiency and application performance. 
VSwapper \cite{vswapper} complements this approach by focusing on the granularity of swap operations. Through its Swap Mapper, VSwapper addresses issues like silent and stale writes by establishing a clear correspondence between disk blocks and guest memory pages. This is achieved via \textit{mmap} system calls, which streamline the swap process and reduce operational overhead, ensuring that the host kernel efficiently manages and names pages.

\subsubsection{Data preservation}

In multi-level memory systems that involve both fast and slow storage spaces, there are two data preservation methods: \textit{inclusive} cache hierarchy and \textit{exclusive} cache hierarchy, as shown in Table \ref{tab:datarelationship}. Oftentimes,  the fast memory entities have a small memory capacity based on volatile memory cells. On the contrary, slow memory entities often have a large capacity on non-volatile memory/storage blocks. 

In an \textit{inclusive} cache hierarchy, the data stored in the fast memory (like RAM or cache) is a subset of the data stored in the slow memory (such as HDDs or SSDs)\cite{book-memorysystems}.
%This means that any piece of data found in the faster memory is also guaranteed to be in the slower memory. 
%The primary purpose of this setup is to ensure data integrity and quick access to frequently used data.
Traditionally, the performance gap between devices in the memory hierarchy is relatively large. Thus, the off-the-shelf often uses inclusive cache hierarchy to reduce end-to-end memory access latency\cite{MSR-linux,memory}. However, as the analysis before (in Sec. \ref{subsec-performance-analysis}), the performance gap has been narrowing with the development of different memory devices. The use of data swap is adequate for better utilization of memory and storage space. In an \textit{exclusive} cache hierarchy, the data stored in the fast and slow memory spaces is distinct\cite{book-memorysystems}. This arrangement entails that the data stored in the fast memory is not duplicated in the slow memory and vice versa. The general approach is to store the most necessary data in faster and often more limited memory space. This method can ensure the maximum utilization of these two types of memory resources. This setup can significantly improve system performance, as it prevents unnecessary duplication and better uses faster memory for critical tasks.

\textbf{Cache optimization:}
Recent works have been focusing on designing data caches between local and far memory. For example, Hybrid$^2$ \cite{hybrid2} designs a memory system architecture that combines a small high-bandwidth 3D-stacked DRAM and a large lower-bandwidth off-chip DRAM. Hydra \cite{lee2022hydra} introduces a configurable resilience mechanism that applies online erasure coding to individual remote memory pages while maintaining high availability. Hydra can be integrated into the existing far memory frameworks such as Infiniswap  \cite{infiniswap}, Leap  \cite{leap} and Remote Regions \cite{remote-regions}, to have better performance.

\textbf{Selective data prefeching: }Prefetching data into local memory before the actual access instruction is often adopted. Efficient prefetching optimizations can happen at the application, system, and hardware levels.  AIFM \cite{aifm} leverages the semantics of data structures to support customized prefetching and caching policies. Fargraph \cite{fargraph} prioritizes data segments based on far memory ratio to minimize transmission overhead. The application's host server uses local memory as a data cache. Fastswap design a prefetch method for higher data access performances \cite{fastswap}. Leap \cite{leap} implements its prefetching algorithm with a data path that segregates application data transfer to far memory. In contrast, Canvas \cite{canvas-nsdi23} schedules packets of each application in an RDMA manner. Furthermore, some works save space and reduce data transfer overhead by compressing cold data  \cite{asplos19software}. HoPP \cite{hopp-huawei} designs a hardware-based memory controller to prefetch sufficient hot pages to OS on a separate data path alongside the normal remote data path. Assise \cite{anderson2020assise} maximizes data locality for all file IO by carrying out IO on process-local, socket-local, and client-local persistent memory. Some works utilize better indexing strategies to accelerate the data fetching. Sherman \cite{wang2022sherman} leverages distributed B+ tree in a DM system to accelerate indexing.

\subsection{Domain-Specific Design}\label{subsec-domain-specific}

% HyFarM \cite{hyfarm} can fit well with existing systems and support excavating the performance and efficiency potential of memory-constrained applications.

\subsubsection{Elastic computing on far memory} 
Elastic applications typically require on-demand dynamic networking. For instance, computing nodes in a disaggregated storage system access the data stored at the storage nodes across the network. A elastic computing system needs to scale automatically according to application demands. For example, KRCore  \cite{wei2022krcore} designs a microsecond-scale control plane on commodity RDMA hardware for elastic computing. The key ideas include virtualizing pre-initialized kernel-space RDMA connections instead of creating one from scratch and retrofitting advanced RDMA dynamic connected transport with static transport for low overhead connection and high networking speed.

\subsubsection{Graph processing on far memory} 
Fargraph \cite{fargraph} analyzes the impact of graph processing workload on disaggregated architecture. Specifically, it reduces the overall data movement through a well-crafted, graph-aware data segment offloading mechanism. In addition, it uses optimal data segment splitting and asynchronous data buffering to achieve graph iteration-friendly far memory access. 
FAMgraph \cite{fam-graph} leverages application-level properties, such as read-only edge data, to efficiently tier data between local and remote memory and prefetch remote data for local computation.

\subsubsection{DNN-supported far memory} 
As the model and dataset for production-scale recommendation systems scale up, the size of the embeddings is approaching the memory capacity limit. Recent studies offload the embedding lookups into SSDs, which target the embedding-dominated recommendation models. RM-SSD \cite{recommendation-ssd} offloads the entire recommendation system into SSD with in-storage computing capability. The proposed SSD-side FPGA solution leverages a low-end FPGA to speed up both the embedding-dominated and MLP-dominated models with high resource efficiency. 
ReCXL\cite{recxl-isca24} is a CXL memory disaggregation system that utilizes near-memory processing for scalable recommendation model training, utilizing data transferring bandwidth optimizations. 
%To meet the strict service level agreement requirements of recommendation systems, the entire set of embeddings needs to be loaded into memory. 
%Limited physical memory constrains the algorithms that can be trained and deployed, posing a severe challenge for deploying advanced inference systems.

\subsubsection{Large-scale model training system}
Due to the huge size of parameters, both training and inference of large language model (LLM) need to be deployed in a distributed manner to improve parallelism. The cost of their hardware and power consumption is exceptionally high. Recent work has investigated high-performance scaling of memory pools and efficient management of hierarchical memory architectures. For example, HET  \cite{hetu-tencent} is a new system framework with a fine-grained cache consistency design that significantly improves the scalability of massive embedding model training. It embraces skewed popularity distributions of embeddings and leverages it to address the communication bottleneck with an embedding cache.% To ensure consistency across the caches, we incorporate a new consistency model into HET design, which provides fine-grained consistency guarantees on a per-embedding basis. Compared to previous work that only allows staleness for read operations, HET also utilizes staleness for write operations. 
Angel-PTM  \cite{Angel-PTM} proposes a productive deep learning system that can train extremely large-scale models pre-training and fine-tuning with hierarchical memory. The key designs of Angel-PTM are fine-grained memory management via page abstraction and a unified scheduling method that coordinates computations, data movements, and communications. Furthermore, Angel-PTM has also designed a lock-free updating mechanism to address SSD I/O bottlenecks.

%% file: Contents/7-Classification.tex
\begin{table}[]\small
\caption{A summary of function on related works. $\checkmark$ means this aspect is carefully designed in the paper,  $\circ$ means this aspect is considered in the paper, $-$ is not considered. "NIC" means Network Interface Card. "CXL" is Computing Express Link. "TCO" means Total Cost of Ownership. "OCM" is Optically Connected Memory. "VM" represents Virtual Machine and "JVM" is Java Virtual Machine. "DL" is Deep Learning and "ML" is Machine Learning. "A/F" represents that the system considers anonymous pages and File-backed pages. "Fa" represents Failure and "Is" represents Isolation.  }
\label{tab:summary of related works}
\vspace{-0.3cm}
 \resizebox{\columnwidth}{!}{%
\begin{tabular}{!{\vrule width1.2pt}c!{\vrule width1.2pt}c|c|c|c|c|c|c|c!{\vrule width1.2pt}c|c|c|c|c!{\vrule width1.2pt}c|c|c|c!{\vrule width1.2pt}c|c|c|c!{\vrule width1.2pt}c|c|c|c|c|c!{\vrule width1.2pt}}%{|c|xxxxxxxx|xxxxx|xxxx|xxxx|}
 \hline

% Old version
% \multicolumn{1}{!{\vrule width1.2pt}c!{\vrule width1.2pt}}{} & \multicolumn{8}{c!{\vrule width1.2pt}}{\textbf{Far Memory Type (Local is DRAM)}} & \multicolumn{5}{c!{\vrule width1.2pt}}{\textbf{System Level}} & \multicolumn{6}{c!{\vrule width1.2pt}}{\textbf{Design Goals}} & \multicolumn{4}{c!{\vrule width1.2pt}}{\textbf{Granularity}} & \multicolumn{4}{c!{\vrule width1.2pt}}{\textbf{Programming}}\\ \hline
% New version
\multicolumn{1}{!{\vrule width1.2pt}c!{\vrule width1.2pt}}{} & \multicolumn{8}{c!{\vrule width1.2pt}}{\textbf{Far Memory Type (Local is DRAM)}} & \multicolumn{5}{c!{\vrule width1.2pt}}{\textbf{Runtime Design Level}} & \multicolumn{4}{c!{\vrule width1.2pt}}{\textbf{Granularity}} & \multicolumn{4}{c!{\vrule width1.2pt}}{\textbf{Programming}}& \multicolumn{6}{c!{\vrule width1.2pt}}{\textbf{Design Metric}}\\ \hline

\multicolumn{1}{!{\vrule width1.2pt}c!{\vrule width1.2pt}}{\begin{tabular}[c]{@{}c@{}}\textbf{Related Works} \\  \\\end{tabular}} & \multicolumn{1}{c|}{\rotatebox{90}{PM (NVM)}} & \multicolumn{1}{c|}{\rotatebox{90}{CXL}} & \multicolumn{1}{c|}{\rotatebox{90}{Network Fabric (NIC)}} & \multicolumn{1}{c|}{\rotatebox{90}{Memory Controller}} & \multicolumn{1}{c|}{\rotatebox{90}{SSD}} & \multicolumn{1}{c|}{\rotatebox{90}{disk}} & \multicolumn{1}{c|}{\rotatebox{90}{GPU}} & \multicolumn{1}{c!{\vrule width1.2pt}}{\rotatebox{90}{Hybrid Memory}} & \multicolumn{1}{c|}{\rotatebox{90}{Hardware-oriented}} & \multicolumn{1}{c|}{\rotatebox{90}{Architecture-oriented}} & \multicolumn{1}{c|}{\rotatebox{90}{Operating System (OS)}} & \multicolumn{1}{c|}{\rotatebox{90}{VM-oriented}} & \multicolumn{1}{c!{\vrule width1.2pt}}{\rotatebox{90}{Service-oriented}} & \multicolumn{1}{c|}{\rotatebox{90}{Cache Line (64B)}} & \multicolumn{1}{c|}{\rotatebox{90}{Page Size (4KB)}} & \multicolumn{1}{c|}{\rotatebox{90}{Large Page (2MB)}}& \multicolumn{1}{c!{\vrule width1.2pt}}{\rotatebox{90}{Object (varied)}}  & \multicolumn{1}{c|}{\rotatebox{90}{Transparency}} & \multicolumn{1}{c|}{\rotatebox{90}{Cache Coherence}} & \multicolumn{1}{c|}{\rotatebox{90}{Customized API usage}} & \multicolumn{1}{c!{\vrule width1.2pt}}{\rotatebox{90}{Parameter Conf.}} & \multicolumn{1}{c|}{\rotatebox{90}{Ownership Cost (TCO)}} & \multicolumn{1}{c|}{\rotatebox{90}{Execution Latency}} & \multicolumn{1}{c|}{\rotatebox{90}{Data/task Throughput \quad}} & \multicolumn{1}{c|}{\rotatebox{90}{Power Consumption \quad}} & \multicolumn{1}{c|}{\rotatebox{90}{Isolation \& Failure  }} & \multicolumn{1}{c!{\vrule width1.2pt}}{\rotatebox{90}{Resource Utilization}}  \\ \hline

Memory Blade'09 \cite{disaggregated-mem-isca09} 
& $-$ & $-$  & $-$ & FGRA & $-$ & $-$ & $-$ &  $-$ 
& $\circ$  & $\checkmark$ & $-$ & $-$ &  $-$

& $-$ & $\checkmark$ &$-$ & $-$  
&$\circ$  &$\checkmark$  & $-$  &$-$
&  $\checkmark$ &  $\checkmark$ & $-$  & $\checkmark$ & $-$ &  $-$ 
\\ \hline

FaRM'14 \cite{farm-nsdi14}
& $-$ & $-$ & RDMA & $-$  & $\circ$ & $-$ & $-$ & $-$ 
& $-$ & $-$ & $-$ & $\checkmark$ & \begin{tabular}[c]{@{}c@{}} Read/Write\end{tabular}

& $-$ & $-$ & $-$ & $\checkmark$
&$-$  &$-$  &$\checkmark$  &$-$
& $-$ & $\checkmark$ & $\checkmark$ & $-$ & $-$ & $-$
\\ \hline

Holistic'15 \cite{holistic2021ccgrid}
& $-$ & $-$ & $-$ & $-$ & $-$ & $-$ & $-$ & NUMA
& $-$ & $-$ & $\checkmark$ & VM &   $-$

& $-$ & $\checkmark$ & $-$ &  $-$
& $\checkmark$   &$-$  &$-$  &$-$
& $-$ & $-$ & $-$ & $-$ & $-$ & $\checkmark$
\\ \hline

HybridSwap'15 \cite{hybridswap}
& $-$ & $-$ & $-$ & $-$  & $\checkmark$  & $-$ & $-$ & $-$
& $-$ & $-$ & $-$ & VM &  $-$

& $-$ & $\checkmark$  & $-$ & $-$
&$\checkmark$ &$-$  &$-$  &$-$
& $-$ & $\checkmark$ & $-$ & $-$ & $-$ & $-$ 
\\ \hline

% XMemPod \cite{xmempod}& $-$ & $-$ & $-$ & $-$ & $-$ & $-$ & $-$ &  & $-$ & $-$ & $-$ & $-$ &  & $-$ & $-$ & $-$ & $-$ & $-$ & $-$ & $-$ & $-$ & $-$ &  &$-$  &$-$  &$-$  &$-$   \\ \hline

LITE'17 \cite{lite}
& $-$ & $-$ & RDMA & $-$  & $-$  & $-$ & $-$ & $-$ 
& $-$ & $-$ & $\checkmark$ & $\checkmark$ & \begin{tabular}[c]{@{}c@{}} Read/Write\end{tabular}

& $-$ & $-$ & $-$ & $\checkmark$
&$-$  &$-$  &$\checkmark$  &$-$  
& $-$ & $\checkmark$ & $\checkmark$ & $-$ & $-$ & $\checkmark$
\\ \hline

Infiniswap'17 \cite{infiniswap}
& $-$ & $-$ & RDMA & $\checkmark$ & $-$ & $-$ & $-$ & $-$
& $-$ & $-$ & $\checkmark$ & $-$ & $-$ 

& $-$ & $\checkmark$ &$-$ & $-$ 
&$\checkmark$  &$-$  &$-$  &$-$  
& $-$ & $\checkmark$ & $-$ & $-$ & Fa &  $-$
\\ \hline

LegoOS'18 \cite{legoos} 
& $\checkmark$ & $-$   & $-$   & $-$ & $\circ$ & $\circ$ & $\circ$  & $-$  
& $-$ & $\circ$  & $\checkmark$ & $\circ$  & $-$ 

& $-$ & $\checkmark$  &$-$ & $-$  
&$\checkmark$  &$-$  &$-$  & $-$  
& $\checkmark$ & $\checkmark$ & $-$ & $-$ & Fa &  $-$ 
\\ \hline

Zombieland'18 \cite{zombieland}
& $-$ & $-$ & RDMA & ACPI & $-$ & $-$ & $-$ & $-$ 
& $-$ & $-$ & $\circ$ & $\checkmark$ & $-$

& $-$ & $\checkmark$ &$-$ & $-$  
&$\checkmark$ &$-$  &$-$  &$-$   
& $-$  & $\checkmark$  & $-$ & $\checkmark$  & $-$ &  $-$ 
\\ \hline

Remote Region'18 \cite{remote-regions}
& $-$ & $-$ & RDMA & $-$  & $\circ$ & $-$ & $-$ & $-$ 
& $-$ & $-$ & $-$ & $\checkmark$ & $-$

& $-$ & $\checkmark$ & $-$ & $\checkmark$
&$-$  &$-$  &$\checkmark$  &$\checkmark$ 
& $-$ & $\checkmark$ & $\checkmark$ & $-$ & Is & $-$
\\ \hline

AR-gRPC'18 \cite{tensorflow-RDMA}  
& $-$ & $-$ & RDMA & $-$  & $-$ & $-$ & $-$ & $-$ 
& $-$ & $-$ & $-$ & $\checkmark$ & Tensorflow

& $-$ & $-$ & $-$ & $\checkmark$
&$-$  &$-$  &$\checkmark$  &$\checkmark$    
& $-$ & $\checkmark$ & $\checkmark$ & $-$ & Is & $-$
\\ \hline

 MC-DLA'18 \cite{micro2018beyond} 
 & $-$ & $-$ & \begin{tabular}[c]{@{}c@{}} PCIe\\ Switch\end{tabular} & $-$ & $-$ & $-$ & $\checkmark$  & $-$
 & $-$ & $\checkmark$  & $-$ & $-$ & \begin{tabular}[c]{@{}c@{}} DL \\ Training\end{tabular}

 & $-$ & $\checkmark$  & $-$ & $-$
 &$-$  &$-$  &$\checkmark$  &$-$  
  & $-$ & $\checkmark$ & $\checkmark$ & $\checkmark$  & $-$ & $-$ 
 \\ \hline
 
String Figure'19 \cite{stringfigure} 
& $-$ & $-$ & $-$ & RTL & $-$ & $-$ & $-$ & HMC  
& $\checkmark$ & $\checkmark$ & $-$ & $-$ &   

&$\checkmark$  & $-$ & $-$ &$-$  
&$-$  &$\checkmark$ &$-$  &$-$
& $-$ & $-$ & $\checkmark$ & $\checkmark$ & $-$ & $-$ 
\\ \hline

Zswap'19 \cite{asplos19software,zswapkernel}
& $-$ & $-$ & $-$ & $-$ & $-$ & $\checkmark$  & $-$ & $-$
& $-$ & $-$ & $\checkmark$ & $-$ & $-$

& $-$ & $\checkmark$ &$-$ & $-$ 
&$\checkmark$  &$-$  &$-$  &$-$ 
& $\checkmark$  & $\checkmark$  & $-$ & $-$ & $-$ &  $-$
\\ \hline

FreeFlow'19 \cite{freeflow}
& $-$ & $-$ & vRDMA & $-$  & $-$ & $-$ & $-$ & $-$ 
& $-$ & $-$ & $-$ & $\checkmark$ & Container

& $-$ & $-$ & $-$ & $\checkmark$
&$-$  &$-$  &$\checkmark$  &$\checkmark$
& $-$ & $\checkmark$ & $\checkmark$ & $-$ & Is & $-$
\\ \hline

Leap'19 \cite{leap} 
& $-$ & $-$ & RDMA & $\checkmark$ & $\checkmark$ & $-$ & $-$ & $-$ 
& $-$ & $-$ & $-$ & $\checkmark$ & $-$

& $-$ & $\checkmark$ & $-$ & $-$
&$-$  &$-$  &$\checkmark$  &$-$  
& $-$ & $\checkmark$ & $\checkmark$ & $-$ & $-$ & $-$
\\ \hline

Clover'20 \cite{pDPM-persistentmemory} 
& $\checkmark$ & $-$ & RDMA & $-$ & $-$ & $-$ & $-$ & $-$
& $-$ & $-$ & $-$ & $\checkmark$ & \begin{tabular}[c]{@{}c@{}}Kv Store\end{tabular}

& $-$ & $-$ &$-$ & $\checkmark$ 
&$-$  &$-$  & $\checkmark$ &$-$  
& $-$ & $\checkmark$ & $\checkmark$ & $\checkmark$ & $-$ &  $-$
\\ \hline

ThymesisFlow'20 \cite{Thymesisflow}
& $-$ & $\circ$  & OpenCAPI & FPGA & $-$ & $-$ & $\circ$  &$-$ 
& $\checkmark$ & $\circ$  & $\checkmark$ & $-$ & $-$ 

& $\circ$  &$\checkmark$ &$-$ & $-$  
&$\checkmark$  &$\circ$  &$-$  &$-$
& $-$ & $-$ & $\checkmark$ & $-$ & $-$ & $\checkmark$
\\ \hline

Fastswap, CFM'20 \cite{fastswap}
& $-$ & $-$ & RDMA & $\checkmark$ & $\checkmark$ & $-$ & $-$ & $-$
& $-$ & $-$ & $\checkmark$ & $\checkmark$ & $-$ 

& $-$ & $\checkmark$ &$-$ & $-$ 
&$\checkmark$  &$-$  &$-$  &$\checkmark$ 
& $\checkmark$ & $\checkmark$ & $\checkmark$ &$-$ & $-$ &  $-$
\\ \hline

Semeru'20 \cite{semeru}
& $-$ & $-$ & RDMA & $\checkmark$ & $\circ$  & $-$ & $-$ &$-$
& $-$ & $-$ & $-$ & JVM & $-$

& $-$ & $\checkmark$  & $-$ &  $\checkmark$ 
&$-$  &$-$  &$\checkmark$   &$-$  
& $-$ & $\checkmark$  & $\checkmark$  & $-$ & $-$ & $\checkmark$ 
\\ \hline

AIFM'20 \cite{aifm}
& $-$ & $-$ & RDMA & $-$ & $-$ & $-$ & $-$ &  $-$
& $-$ & $-$ & $-$ & $\checkmark$ & $-$ 

& $-$ & $-$ & $-$ & $\checkmark$
&$-$  &$-$  &$\checkmark$  &$-$ 
& $-$ & $\checkmark$ & $\checkmark$ & $-$ & $-$ & $-$ 
\\ \hline

Swap Advisor'20\cite{swapadvisor}
& $-$ & $-$ & NVLink & $-$ & $\checkmark$  & $-$ & $\checkmark$ & $-$
& $-$ & $-$ & $-$ & $\checkmark$ &\begin{tabular}[c]{@{}c@{}} DL \end{tabular}
 
& $-$ & $\checkmark$ & $-$ & $-$ 
&$\checkmark$ &$-$  &$-$  &$\checkmark$
& $-$ & $\checkmark$  & $\checkmark$  & $-$ & $-$ & $-$
\\ \hline

 Buddy Compression'20 \cite{buddycompression} 
& $-$ & $-$ & NVLink & $-$  & $\checkmark$ & $-$ & $\checkmark$  & $-$
& $-$ & $-$ & $-$ &$\checkmark$  & $-$

& $-$ & $\checkmark$ & $-$ & $-$
&$-$  &$\checkmark$  &$-$  &$-$
& $-$ & $\checkmark$ & $\checkmark$  & $-$ & $-$ & $-$
\\ \hline

 Griffin'20 \cite{griffin}
 & $-$ & $-$ & PCIe & IOMMU & $-$ & $-$ & $\checkmark$ & NUMA
 & $\checkmark$  & $\checkmark$  & $-$ & $-$ & $-$ 
 
 & $-$ & $\checkmark$  & $-$ & $-$ 
 &$\checkmark$   &$-$  &$-$  &$-$ 
  & $-$ & $\checkmark$  & $-$ & $-$ & $-$ & $-$
 \\ \hline

Hybrid${^{2}}$'20 \cite{hybrid2} 
& $-$ & $-$ & DIMM & $-$ & $-$ & $-$ & $-$ & \begin{tabular}[c]{@{}c@{}} 3D-stacked \\ DRAM\end{tabular}
& $\checkmark$ & $\checkmark$ & $-$ & $-$ & $-$

& $-$ & $\checkmark$ & $-$ & $\checkmark$ 
&$-$  &$-$  &$\checkmark$  & $\checkmark$ 
& $\checkmark$ & $\checkmark$ & $\checkmark$ & $-$ & $-$ & $-$ 
\\ \hline

MIND'21 \cite{mind}
& $-$ & $-$ & Switch & ASIC & $-$ & $-$ & $-$ & $-$
&$\circ$  &$\checkmark$ & $-$ & $-$ & $-$

& $-$ & $\checkmark$ &$-$ & $-$ 
&$\checkmark$  &$-$  &$-$  &$-$
& $-$ & $\checkmark$ & $-$ & $-$ & Fa &  $-$
\\ \hline

XMemPod'21 \cite{icdcs19, FastSwap-github,xmempod}
& $-$ & $-$ & RDMA & $\checkmark$ & $\checkmark$ & $-$ & $-$ & $-$
& $-$ & $-$ & $\checkmark$ & VM & $-$ 

& $-$ & $\checkmark$ &$-$ & $-$ 
&$\checkmark$  &$-$  &$-$  &$-$ 
& $-$ & $\checkmark$ & $-$ & $-$ & Is &  $-$
\\ \hline

Kona'21 \cite{rethinking-asplos21}
& $-$ & $\checkmark$ & RDMA & FPGA & $-$ & $-$ & $-$ & $-$  
& $-$ & $\checkmark$ & $-$ & $\checkmark$ & $-$  

& $\checkmark$  & $-$ & $-$ & $-$    
&$\checkmark$  &$\checkmark$  & $\checkmark$  &$-$
& $-$ & $\checkmark$ & $-$ & $-$ & Fa & $-$
\\ \hline

ZeRO-Infinity'21 \cite{Zero-infinity} 
& $-$ & $-$ & NVLink &  $-$ & $\checkmark$ & $-$ & $\checkmark$  & $-$
& $-$ & $-$ & $-$ &$\checkmark$  & \begin{tabular}[c]{@{}c@{}} DL \end{tabular}

& $-$ & $\checkmark$ & $-$ & $-$
&$-$  &$\checkmark$  &$-$  &$-$
& $-$ & $\checkmark$ & $\checkmark$  & $-$ & $-$ & $-$
\\ \hline
 
% LaySA'21 \cite{zhao2021bridging} 
% & $-$ & $-$ & $-$ & $-$ & $-$ & $-$ & $\checkmark$ & $-$
% & $-$ & $-$ & $-$ & $-$ & \begin{tabular}[c]{@{}c@{}} DL \\ Training\end{tabular}

& $-$ & $\checkmark$  & $-$ & $-$
&$\checkmark$   &$-$  &$-$  &$-$
& $-$ & $\checkmark$  & $\checkmark$  & $-$ & $-$ & $-$ 
\\ \hline

HET'21 \cite{hetu-tencent} 
& $-$ & $-$ & $-$ & $-$  & $-$ & $\checkmark$ & $\checkmark$  & $-$
& $-$ & $-$ & $-$ &$\checkmark$  & \begin{tabular}[c]{@{}c@{}} ML \\ Training\end{tabular}

& $-$ & $\checkmark$ & $-$ & $-$
&$-$  &$\checkmark$  &$-$  &$-$ 
& $\checkmark$ & $\checkmark$ & $-$  & $-$ & $-$ & $-$
\\ \hline

Memtrade'21 \cite{memtrade}
& $-$ & $-$ & $-$ & $-$ & $\checkmark$ & $\checkmark$ & $-$ & $-$
& $-$ & $-$ & $-$ & VM & $-$

& $-$ & $\checkmark$ & $-$ & $-$
&$-$  &$-$  &$\checkmark$ &$-$
& $\checkmark$ & $\checkmark$ & $-$ & $-$ & $-$ & $\checkmark$
\\ \hline

TMO'22 \cite{tmo}
& $-$ & $-$ & $-$ & $-$ & $\checkmark$ & $\checkmark$ & $-$ & $-$
& $-$ & $-$ & $\checkmark$ & $-$ & $-$

& $-$ & A/F & $-$ & $-$
&$\checkmark$  &$-$  &$-$  &$\checkmark$
& $\checkmark$ & $\checkmark$ & $\checkmark$ & $-$ & $-$ & $-$
\\ \hline

Sherman'22 \cite{wang2022sherman}
& $\circ$ & $-$ & RDMA & $-$ & $-$ & $-$ & $-$ & $-$ 
& $-$ & $-$ & $-$ & $\checkmark$ & \begin{tabular}[c]{@{}c@{}}Read/Write\end{tabular}

& $-$ & $-$ &$-$ & $\checkmark$  
&$-$  &$-$  &$\checkmark$  &$-$ 
& $-$ & $\checkmark$ & $\checkmark$ & $-$ & $-$ &  $-$ 
\\ \hline

Teleport'22\cite{teleport-sigmod}
& $-$ & $-$   & RDMA    & $-$ & $\circ$ & $\circ$ & $\circ$  & $-$  
& $-$ & $\circ$  & $\checkmark$ & $\circ$  & $-$ 

& $-$ & $\checkmark$  &$-$ & $-$  
&$-$  &$\checkmark$  &$\checkmark$  & $-$
& $\checkmark$ & $\checkmark$ & $-$ & $-$ & $-$  &  $-$ 
\\ \hline

Jiffy'22 \cite{jiffy-eurosys22}
& $-$ & $-$ & RDMA &$\checkmark$ &$\checkmark$& $-$ & $-$ & $-$
& $-$ & $-$ & $-$ & $\checkmark$ & Serverless

& $-$ & $-$ & $-$ & $\checkmark$
&$-$  &$-$  &$\checkmark$   &$\checkmark$
& $\checkmark$ & $\checkmark$ & $-$ & $-$ & $-$ & $\checkmark$ 
\\ \hline

Clio'22 \cite{clio}
& $-$ & $-$ & RDMA & FPGA  & $-$ & $-$ & $-$ & $-$ 
& $\circ$ & $\checkmark$ & $-$ & $-$ & $-$  

& $-$ & $\circ$ & $-$ & $\checkmark$
&$-$  &$-$  &$\checkmark$ &$-$  
& $-$ & $\checkmark$ & $\checkmark$ & $\checkmark$ & $-$ & $-$
\\ \hline

OCM'22 \cite{optically-jpdc2022}
& $-$ & $-$  & Optical & $-$ & $-$ & $-$ & $-$ & OCM
& $\checkmark$ & $\checkmark$ & $-$ & $-$ & $-$

& $\checkmark$ & $-$ &$-$ & $-$  
&$\checkmark$ &$-$  &$-$  &$-$  
& $-$ & $\checkmark$& $\checkmark$ & $\checkmark$ & $-$ &  $-$ 
\\ \hline

BEACON'22 \cite{beacon-micro22}
& $-$ & $\checkmark$ & $-$ & $-$ & $-$ & $-$ & $-$ & $\circ$ 
& $\circ$ & $\checkmark$ & $-$ & $-$ & \begin{tabular}[c]{@{}c@{}}Genome \\ Analysis\end{tabular}

& $\checkmark$ & $-$ &$-$ & $-$  
& $-$  &$-$  &$\checkmark$  & $\checkmark$  
& $-$ & $\checkmark$ & $-$ & $\circ$ & $-$ &  $-$ 
\\ \hline

MemLiner'22 \cite{memliner}
& $-$ & $-$ & RDMA & $\checkmark$  & $-$ & $-$ & $-$ & $-$  
& $-$ & $-$ & $-$ & JVM & $-$ 

& $-$ & $-$ & $-$ &  $\checkmark$  
&$-$  &$\circ$  & $\checkmark$   &$-$ 
& $\checkmark$  & $\checkmark$  & $-$ & $-$ & $-$ & $-$ 
\\ \hline

Mitosis'22\cite{mitosis-serverless}
& $-$ & $-$ & RDMA & $-$  & $-$ & $-$ & $-$ & $-$ 
& $-$ & $-$ & $\checkmark$ & $-$ & Serverless

& $-$ & $-$ & $-$ & $\checkmark$
&$-$  &$-$  &$\checkmark$  &$\checkmark$
& $-$ & $\checkmark$ & $\checkmark$ & $-$ &Is & $-$
\\ \hline

Medes'22 \cite{saxena2022memory}
& $-$ & $-$ & RDMA & $-$  & $-$ & $-$ & $-$ & $-$ 
& $-$ & $-$ & $-$ & $\checkmark$ & Serverless

& $-$ & $\checkmark$ & $-$ & $\circ$
&$-$  &$-$  &$\checkmark$  &$\checkmark$ 
& $\checkmark$ & $\checkmark$ & $\checkmark$ & $-$ &Is & $-$
\\ \hline

Fargraph'22 \cite{fargraph, fargraphplus} 
& $-$ & $-$ & RDMA & $\checkmark$  & $\checkmark$ & $-$ & $-$ & 
& $-$ & $-$ & $-$ & $\checkmark$ & Graph

& $-$ & $-$ & $-$ & $\checkmark$ 
&$-$  &$-$  & $\checkmark$  &$-$   
& $\checkmark$ & $\checkmark$ & $\checkmark$ & $-$ & $-$ & $-$
\\ \hline

FAMgraph'22\cite{fam-graph}
& $-$ & $-$ & RDMA & $\checkmark$  & $\checkmark$ & $-$ & $-$ & 
& $-$ & $-$ & $-$ & $\checkmark$ & Graph

& $-$ & $-$ & $-$ & $\checkmark$ 
&$-$  &$-$  & $\checkmark$  &$-$ 
& $\checkmark$ & $\checkmark$ & $\checkmark$ & $-$ & $-$ & $-$
\\ \hline

KRCore'22 \cite{wei2022krcore}
& $-$ & $-$ & RDMA & $-$ & $-$ & $-$ & $-$ & $-$
& $-$ & $-$ & $\checkmark$& $-$ & Elastic

& $-$ & $-$ & $-$ & $\checkmark$
&$-$  &$-$  &$\checkmark$  &$-$   
& $-$ & $\checkmark$  & $\checkmark$ & $-$ & $-$ & $-$ 
\\ \hline

RM-SSD'22\cite{recommendation-ssd} 
& $-$ & $-$ & FPGA & $\checkmark$  & $-$ & $-$ & $-$ &$-$   
& $\checkmark$ & $\circ$ & $-$ & $-$ & $-$

& $-$ & $\checkmark$  & $-$ & $-$
&$-$  &$-$  & $\checkmark$  &$-$
& $-$ & $\checkmark$ & $\checkmark$ & $-$ & $-$ & $-$ 
\\ \hline

COARSE'22 \cite{coarse-hpca22} 
& $-$ & $-$ & CCI & FPGA & $-$ & $-$ & $\checkmark$  & $-$
& $-$ & $-$ & $-$ & $-$ & \begin{tabular}[c]{@{}c@{}} DL \\ Training\end{tabular}

& $-$ & $\checkmark$ & $-$ & $-$
&$-$  &$\checkmark$  &$-$  &$-$ 
& $-$ & $\checkmark$ & $\checkmark$  & $-$ & $-$ & $-$
\\ \hline

HyFarM'22 \cite{hyfarm-iccd}
& $-$ & $-$ & RDMA & $-$ & $\checkmark$ & $\checkmark$ & $-$ & $-$
& $-$ & $-$ & $-$ & $\checkmark$  & $-$

& $-$ & A/F & $-$ & $-$
& $\checkmark$ &$-$  &$-$  &$\checkmark$
& $\checkmark$  & $\checkmark$  & $-$ & $-$ & $-$ & $\checkmark$ 
\\ \hline

Canvas'23 \cite{canvas-nsdi23}
& $-$ & $-$ & RDMA & $-$ & $-$ & $-$ & $-$ & $-$
& $-$ & $-$ & $\checkmark$  & $\checkmark$  & $-$

& $-$ & A/F  & $-$ &  $-$
&$\checkmark$  &$-$  &$-$  &$\checkmark$ 
& $-$ & $\checkmark$ & $-$ & $-$ & Is & $-$
\\ \hline

BAM'23 \cite{bam}
& $-$ & $-$ & $-$ &  $-$ & $\checkmark$ & $-$ & $\checkmark$  & $-$
& $-$ & $-$ & $-$ &$\checkmark$  & $-$

& $-$ & $\checkmark$ & $-$ & $-$
&$-$  &$\checkmark$  &$-$  &$-$
& $-$ & $\checkmark$ & $\checkmark$  & $-$ & $-$ & $-$
\\ \hline

GIDS'23 \cite{GIDS}
& $-$ & $-$ & $-$ &  $-$ & $\checkmark$ & $-$ & $\checkmark$  & $-$
& $-$ & $-$ & $-$ &$\checkmark$  & $-$

& $-$ & $\checkmark$ & $-$ & $-$
&$-$  &$\checkmark$  &$-$  &$-$   
& $-$ & $\checkmark$ & $\checkmark$  & $-$ & $-$ & $-$
\\ \hline

Helios'23 \cite{helios} 
& $-$ & $-$ & $-$ &  $-$ & $\checkmark$ & $-$ & $\checkmark$  & $-$
& $-$ & $-$ & $-$ &$\checkmark$  & $-$

& $-$ & $\checkmark$ & $-$ & $-$
&$-$  &$\checkmark$  &$-$  &$-$   
& $-$ & $\checkmark$ & $\checkmark$  & $-$ & $-$ & $-$
\\ \hline

Angel-PTM \cite{Angel-PTM} 
& $-$ & $-$ & $-$ & $-$  & $\checkmark$ & $-$ & $\checkmark$  & HBM
& $-$ & $-$ & $-$ &$\checkmark$  & \begin{tabular}[c]{@{}c@{}} LLM \\ Training\end{tabular}

& $-$ & $\checkmark$ & 4MB & $-$
&$-$  &$\checkmark$  &$-$  &$-$
& $\checkmark$  & $\checkmark$ & $\checkmark$  & $-$ & $-$ & $-$
\\ \hline

Pond'23 \cite{pond-cxl}
& $-$ & $\checkmark$ & \begin{tabular}[c]{@{}c@{}} CXL \\ Switch\end{tabular} & $-$ & $-$ & $-$ & $-$  & NUMA
& $\checkmark$  & $-$ & $-$ & VM  & $-$

& $-$ &$\checkmark$ &  $-$ &  $-$
& $\checkmark$  & $\circ$  &$-$  &$-$   
& $\checkmark$& $\checkmark$& $-$ & $-$ & $-$ &  $\checkmark$ 
\\ \hline

xDM'24 \cite{wang-xdm-sc24}
& $-$ & $\checkmark$ & \begin{tabular}[c]{@{}c@{}} RDMA\end{tabular} & $-$ & $\checkmark$  & $\checkmark$  & $-$  & NUMA
& $-$  & $\checkmark$  & $\checkmark$  & VM  & $-$

& $-$ &$\checkmark$ &  2MB &  $-$
& $\checkmark$  & $\circ$  &$-$  & $\checkmark$  
& $\checkmark$& $\checkmark$& $-$ & $-$ & Is &  $\checkmark$ 
\\ \hline

GMT'24 \cite{gmt-asplos24}
& $-$ & $-$ & $-$ & $-$ & $\checkmark$  & $\checkmark$  & $\checkmark$  & $-$
& $\checkmark$  & $\checkmark$  & $-$  & $-$ & $-$

& $\checkmark$ & $\checkmark$ & $-$ &  $-$
& $\checkmark$  & $\circ$  &$-$  & $\checkmark$  
& $-$ & $\checkmark$& $-$ & $-$ & $-$ & $-$ 
\\ \hline

ReCXL'24 \cite{recxl-isca24}
& $-$ & $\checkmark$ & $-$ & PNM & $-$ & $-$  & $\checkmark$   & $\checkmark$ 
& $\checkmark$  & $\checkmark$  & $-$  & $-$  & \begin{tabular}[c]{@{}c@{}} AI \\ training\end{tabular}

& $\checkmark$ & $\checkmark$ & $-$ &  $-$
& $\checkmark$  & $\circ$  &$-$  & $\checkmark$  
& $-$ & $\checkmark$& $-$ & $\checkmark$ & $-$ & $-$ 
\\ \hline

\end{tabular}%
}
\end{table}

% Beldi \cite{zhang2020fault} 
% & $-$ & $-$ & $-$ & $-$ & $-$ & $-$ & $-$ &  
% & $-$ & $-$ & $-$ & $-$ & 
% & $-$ & $-$ & $-$ & $-$ & $-$ & $-$ 
% & $-$ & $-$ & $-$ & 
% &$-$  &$-$  &$-$  &$-$   \\ \hline

\section{Summary of Cross-layer DM Works}\label{sec-classification}
%\textbf{Memory Hardware Architecture:}
In this section, we give an overall summary about representative disaggregated memory works with cross-layer designs, i.e. architecture level, system level, and application level. We summarize the far memory types (architecture level), runtime design level (system level), design considerations including data granularity, programming difficulty, and design metric (application level).

\textbf{Far memory type:}
The second column in Table ~\ref{tab:summary of related works} shows far memory types used in recent works. From the perspective of hardware, most works choose DRAM as local memory. There are plenty of works supporting NIC-based far memory. A subset of studies explore alternative NIC designs except RDMA NIC, including PCIe Switch, OpenCAPI, NVLink, DIMM, Optical, FPGA, CCI, and CXL Switch. 
Particularly serving for GPU-like accelerators, GPU far memory is becoming a new trend. Related works have supported the heterogeneous memory types of NUMA, HMC, 3D-stacked DRAM, OCM, and HBM, which can also viewed as far memory. With the advancement of storage devices such as NVM, PM, SSD, and disk, storage-based far memory has started attracting interest. Furthermore, there has been a gradual increase in research supporting CXL. The design of memory controllers for far memory pool management is increasing, such as FGRA, ACPI, RTL, FPGA, IOMMU, and ASIC.
% TODO: hybrid 新？PM traditional？

% PM(NVM): Only LegoOS'18 \cite{legoos} and Clover'20 \cite{pDPM-persistentmemory} carefully design Persistent Memory(Non-Volatile Memory) in the paper, and Sherman'22 \cite{wang2022sherman} considers PM in the paper. The other related works don't consider PM.
% CXL: Only three works carefully design CXL in their paper, which are Kona'21 \cite{rethinking-asplos21}, BEACON'22 \cite{beacon-micro22}, and Pond'23 \cite{pond-asplos23}. ThymesisFlow'20 \cite{thymesisflow} also consider CXL in the paper.
% Network Fabric(NIC): The majority of related works carefully design NIC, among which most of designed NICs integrate RDMA in their paper, 
% Memory Controller: A few related works in memory controllers design include FGRA, ACPI, RTL, FPGA, IOMMU, ASIC, and FPGA.
% SSD: Quite a few related works in memory architecture design include SSD.
% disk and GPU: Only a few related works in memory architecture design include SSD and GPU. Only HET'21 \cite{hetu-tencent} carefully design both GPU and SSD in memory architecture, and LegoOS'18 \cite{legoos} consider both in their work.
% Hybrid Memory: Few related works use hybrid memory, specifically: NUMA, HMC, 3D-stacked DRAM, and OCM.  

%\textbf{System-Level Considerations:}
\textbf{Runtime design level:}
We list 5 design considerations in the third column of Table ~\ref{tab:summary of related works}, including specific hardware-oriented system designs  (Sec.  \ref{subsec-memorydevice} and \ref{subsec-performance-analysis}), architecture designs (Sec. \ref{subsec-cluster-level}), OS-based adaptions (Sec.  \ref{subsec-os-design}), VM-centric designs on virtualized environments (Sec. \ref{subsec-virtualized}), and services-aware designs (Sec. \ref{subsec-domain-specific}). 
Most works implement optimization methods in virtualized environments, including virtual machines and Java virtual machines.  Nearly a third of works design for a specific service, including optimized systems on data read/write access, tensorflow, deep learning training, container-based optimization, key-value store, serverless, genome analysis, graph processing, and elastic computing services. 

%As technologies evolves, the focus of services designed in system level has been changing. Early systems were built for basic tasks like reading and writing data or managing software containers. But now, they're being made to handle more complex jobs, such as training for Deep Learning and Large Language Models.

%A few works are oriented towards hardware, architecture and operating systems as listed in Table ~\ref{tab:summary of related works} respectively.
%\textbf{Runtime and Service Considerations:}

\textbf{Data granularity:}
We list the supported data granularity in the fourth columns of Table ~\ref{tab:summary of related works}. 
There are three general types of fixed-size data granularity, the cacheline size in 64B, the page size in 4KB, and the large page size in 2MB. 
Most of the works support  4KB page size, with more papers tents to use a large page size as 4MB in Angel-PTM \cite{Angel-PTM}. A few works enable 64-byte cache line architecture in data transfer: String Figure \cite{stringfigure}, Kona \cite{rethinking-asplos21}, OCM \cite{optically-jpdc2022}, and BEACON \cite{beacon-micro22}. Many works use a flexible and varied data block size. 

%and we summarized them as data objects
\textbf{Program programming difficulty:}
As shown in the fifth columns of Table ~\ref{tab:summary of related works}, when analyzing the programming difficulties of programs in the related works, we consider four parts: programming transparency, cache coherence, customized API usage, and parameter configurations. We observe that quite a few works are designed to be transparent and enable configuration parameters. Recent related works have started to design cache coherence protocols, especially on CXL-based memory. Most application-oriented works design customized API usage to utilize RDMA, or far memory calls for user-friendly programming.

%\textbf{Design Goals:}
\textbf{System design Metric:}
We list typical design metrics in the sixth column of Table ~\ref{tab:summary of related works}. In general, the design goals of the listed works aim at better functionality and higher performance. The detailed categories include the total cost of ownership (TCO), execution latency, data/task throughput, power consumption, isolation and failure, and resource utilization.
The majority of systems are designed to reduce execution latency and improve data or task throughput. Some works mainly focus on reducing total cost of ownership (TCO) and power consumption by using efficient hardware or design power management software. Systems can also add secure-aware functions by realizing resource isolation (marked as Is) and handling failures (marked as Fa).

%In summary, we give a detailed classification of existing remarkable disaggregated memory systems and their features. This table reflects past and future design trends, which can inspire more work contributing to the family of disaggregated memory systems. 

%% file: Contents/8-Futurework.tex
\section{Future Work}\label{sec-futurework}

%Future work on DM will focus on high performance and low cost. Devices will be designed for high data bandwidth and low memory access latency. The lightweight and flexible software will support more functions and higher task throughput. 

\subsection{Emerging technology of disaggregated memory dystem}
The design of the DM system requires lower-power and high-performance hardware devices, as well as support for higher concurrency and larger-scale applications.

\textbf{Hardware design for DM: } Existing works try to design and develop memory hardware with the ability of near-data processing, direct connection, hot swap/plugging, data consistency, and security guarantee. CXL-supported memory still needs further design on cross-node and cross-rack memory infrastructure. This will increasingly rely on the design of higher-speed network transfers and more efficient network bandwidth allocation.
    %A key trend for hardware design is to utilize DPU to offload the memory control modules on DPU and access far memory with high throughput while keeping low power consumption. 

% \subsubsection{Operating System and Runtime Design for DM}
% \textit{Resource Abstraction and Isolation: }
% \textit{Heterogeneous resource management: }

% \textit{High throughput FM paths: }
% \textit{Light-weight FM access: }

\textbf{DM for cloud services:} Adapting far memory to platform as a service (PaaS), software as a service SaaS, function as a service (FaaS), and model as a service (MaaS) is essential. This will also lead to many promising research problems and novel solutions on virtualization environment design for services. DM system design needs to reduce the overhead of far memory access in virtualized environments.

\textbf{ Memory pool for xPUs:} Today, the computation center is gradually migrating from CPUs to GPUs, FPGA, and processing-in-memory (PIM) accelerators, also named as x processing units (xPUs). Existing work uses distributed shared memory to design systems, and there is still a blue ocean of collaboration for heterogeneous compute resources with memory pools. Providing GPUs with larger memory storage space and higher-speed data access will lead to explosive performance gains.

\textbf{DM for micro datacenter:} In the era of micro datacenter, i.e. edge datacenter,  edge devices tend to have more restricted memory and storage space and thus require fine-grained data offload design. Micro datacenters can provide higher computing power and a more flexible allocation of computing resources than end devices. Deploying disaggregated architecture in micro data centers will improve overall resource efficiency.

\textbf{High memory bandwidth and utilization: } Multiple modality natural networks have become popular in today's applications. It is essential to build highly parallel and high-bandwidth GPU remote access pathways. Designing high-performance GPU remote memory access patterns can be a main trend in future work. Further, improving multiple data transfer channels is also a good solution with good coordination of complex memory hierarchies with high parallelism.

\subsection{ Disaggregated Memory for Emerging Applications}
The explosive growth of AI model size inevitably increases the computation and memory costs, requiring large amount of GPUs. Managing memory resources is especially important for the pre-training and fine-tuning retraining process of large models, which needs to ensure the training latency and effectiveness.

\textbf{Design for large foundation model: } In sparse LFM scenarios, the number of longest tokens determines the matrix's length and width, so many zeros are in the matrix. This will occupy a lot of valuable GPU memory resources. One solution is to utilize multi-level memory to offload cold data to slow storage devices to utilize GPU memory space more efficiently. Using a polarized de-redundancy approach, the flexible memory extension method for LFM can accommodate more important computational data.

% \textbf{Design for incremental learning: } Incremental learning can improve the existing model incrementally with new samples continuously added to AI models, which is widely used today. However, incremental learning has long faced the problem of "catastrophic forgetting". The model will forget the previously learned knowledge and overfit to the new data.   Establishing high-speed far memory channels with reasonable caching mechanisms can ensure fast data read-in and write-out.

%It is interesting to research on how incremental learning under the multimodal large model task will affect its consumption of computational memory resources.
% \red{to be extended}

\textbf{Design for multi-modal neural network:} Disaggregated memory provides significant performance improvements, higher memory utilization efficiency, and robust model training capabilities for multi-modal neural networks by separating memory resources from processing units. It allows one to handle large-scale datasets from different data sources, supporting more complex network structures while reducing bottlenecks caused by memory limitations. It can greatly enhance the scalability and flexibility of the model. %Disaggregated memory technology is crucial in improving multi-modal neural networks' processing power and efficiency, especially suitable for scenarios that require large amounts of data processing and high-performance computing.

\textbf{Design for metaverse and cloud gaming:}  Disaggregated memory technology significantly enhances performance and optimizes metaverse and cloud gaming resources. By enabling seamless, cross-server sharing of memory resources within data centers, this technology allows metaverse and cloud gaming platforms to handle large-scale user data and complex graphic rendering tasks with greater efficiency. %Not only does this help reduce latency and increase response times, thereby improving the user experience, but it also effectively minimizes hardware resource waste and operational costs. As a result, these platforms can support more concurrent users, facilitating richer interactions and more immersive virtual experiences.

\textbf{Design for embodied artificial intelligence:} DM holds transformative potential for embodied artificial intelligence, enhancing systems like autonomous vehicles and interactive robots by enabling dynamic scalability, improved memory utilization, and reduced latency. The technology allows AI systems to dynamically adjust and allocate memory resources across multiple nodes, which is essential for efficiently handling large, unpredictable data streams and complex computations.% Consequently, disaggregated memory boosts the performance and adaptability of embodied AI applications and drives cost efficiency by optimizing memory usage across diverse and dynamic environments.

% \textbf{Quantum Computing on DM:}

\textbf{Design for scientific computing:} Disaggregated memory significantly enhances the efficiency and flexibility of high-performance computing systems in scientific computing, such as genomics, by allowing dynamic allocation of memory resources independent of compute nodes. This adaptability improves data load management effectiveness, reduces memory bottlenecks, and optimizes the overall system performance and cost efficiency. %Consequently, disaggregated memory enables more complex scientific analyses and considerable dataset processing, which are critical for advancing research capabilities in computationally intensive fields.